\documentclass[report]{aa}

\usepackage{graphicx}
\usepackage{xcolor}
\usepackage{txfonts}
\usepackage{natbib}
\usepackage{float}

\newcommand{\chg}[1]{#1}

\begin{document}

\title{Photodissociation of interstellar N$_2$}


\author{Xiaohu Li\inst{1}
        \and Alan N.\ Heays\inst{1,2,3}
        \and Ruud Visser\inst{4}
        \and Wim Ubachs\inst{3}
        \and Brenton R.\ Lewis\inst{5}
        \and Stephen T.\ Gibson\inst{5}
          \and Ewine F. van Dishoeck\inst{1,6}
        }

        \institute{$^1$ Leiden Observatory, Leiden University,
          P.O. Box 9513, 2300 RA Leiden, The Netherlands\\
          $^2$ Department of Physics, Wellesley College, Wellesley, MA 02181, USA \\
          $^3$ Department of Physics and Astronomy, LaserLaB, VU
          University, de Boelelaan 1081, 1081 HV Amsterdam, The Netherlands\\
          $^4$ Department of Astronomy, University of Michigan, 500 Church Street, Ann Arbor, MI 48109-1042, USA \\
          $^5$ Research School of Physics and Engineering, The Australian National University, Canberra, ACT 0200, Australia \\
          $^6$ Max-Planck Institut f\"ur Extraterrestrische Physik
          (MPE), Giessenbachstr.\ 1, 85748 Garching, Germany \\
          \email{li@strw.leidenuniv.nl}}

\date{\textbf{Draft -- \today}}

\titlerunning{Photodissociation of interstellar N$_2$}
\authorrunning{Li et al.}


\def\placefigpotentials{     
\begin{figure}[htb]
\centering
\includegraphics[angle=0,width=1\hsize]{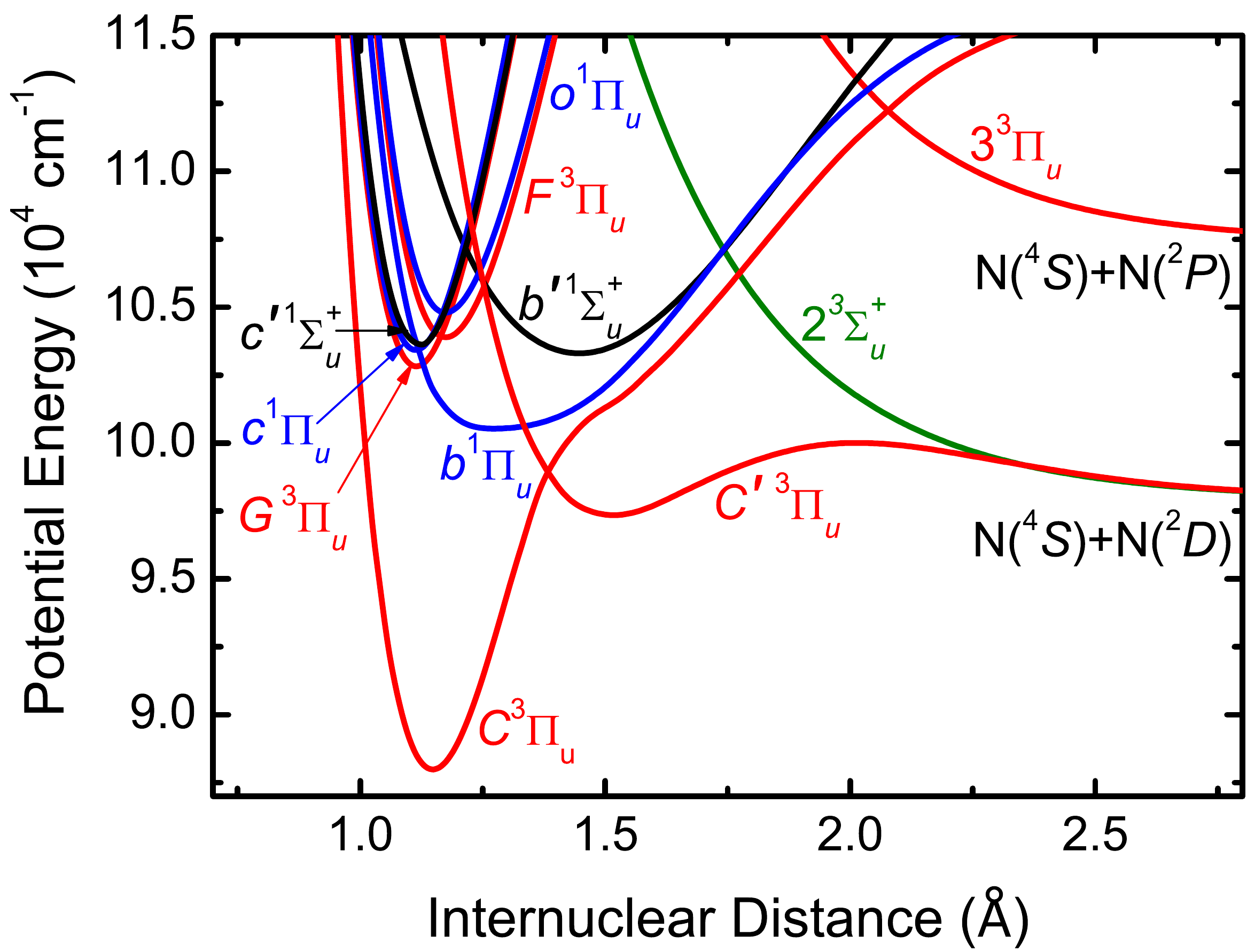}
\caption{Diabatic-basis potential-energy curves for electronic states
  of N$_2$ relevant to interstellar photodissociation.  Blue curves:
  $^1\Pi_u$ states. Black curves: $^1\Sigma_u^+$ states. Red curves:
  $^3\Pi_u$ states. Green curve: $2\,{^3\Sigma_u^+}$ state.  The
  energy scale is referenced to the $v=0,J=0$ level of the
  $X\,{^1\Sigma_g^+}$ ground state (not shown). The lowest
  dissociation limit, N$(^4S)+$N$(^4S)$ at $\sim78\,715$~cm$^{-1}$
  (9.76 ev), is beyond the scale of the figure.  The H ionization
  potential of 13.6 eV provides an upper limit to the interstellar
  radiation field and corresponds to 109691 cm$^{-1}$.}
\label{fig:N2potentials}
\end{figure}
}   

\def\placefigoverview{
\begin{figure*}[htb] \centering
    \includegraphics[angle=0,width=1\hsize,height=6.5cm]{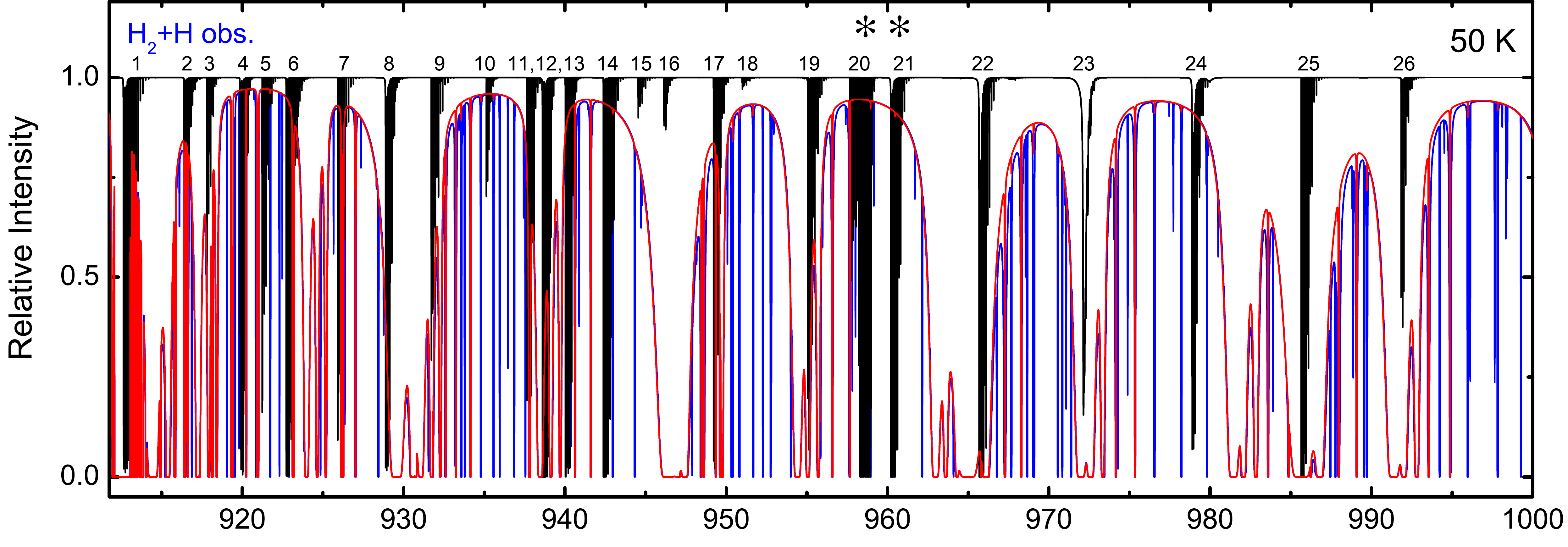}
    \includegraphics[angle=0,width=1\hsize,height=6.5cm]{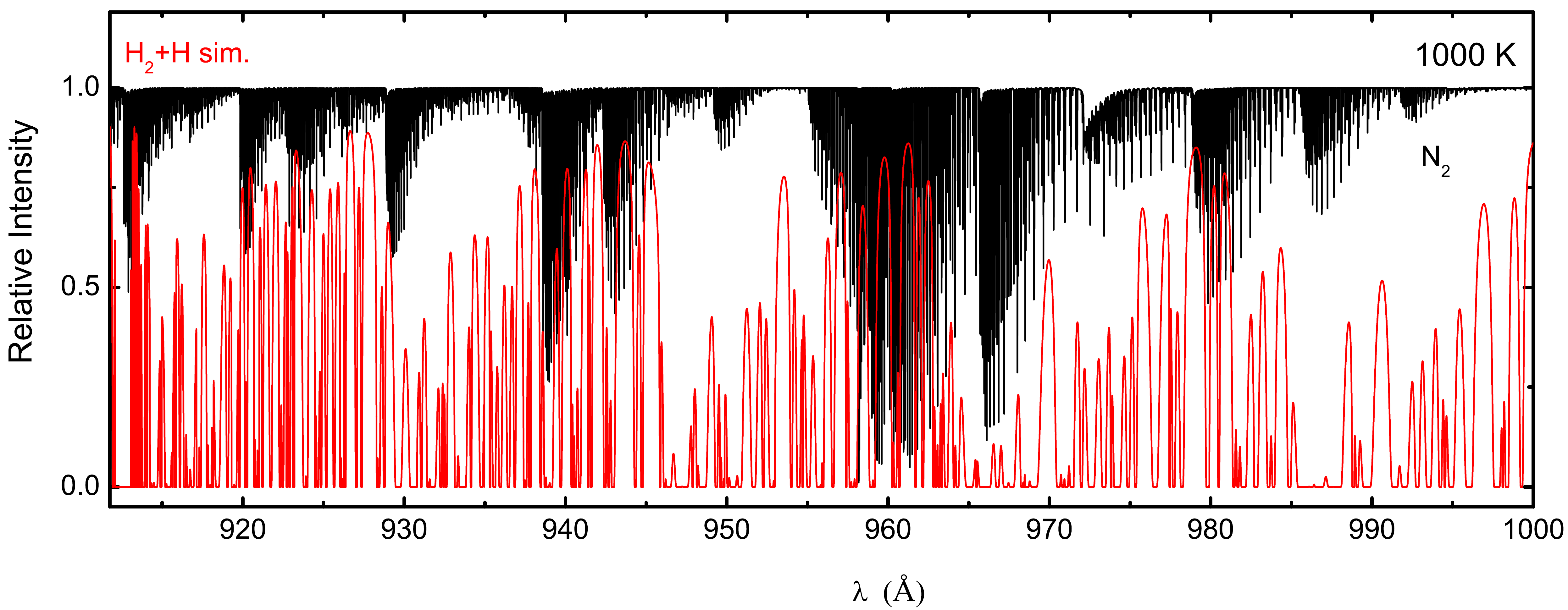}
    \caption{Simulated absorption spectra for N$_2$ (black) and
      ${\rm{}H}_2+{\rm{}H}$ (red) in the wavelength range
      912--1000\,\AA{} assuming thermal excitation temperatures of 50
      (top) and 1000~K (bottom). The column density of N$_2$ is
      $10^{15}$\,cm$^{-2}$ and values for H$_2$ and H are taken to be
      half of the observed column densities in the well-studied
      diffuse cloud toward $\zeta$ Oph, as is appropriate for the
      center of the cloud: $N$(H$_2$)=$2.1\times 10^{20}$ and
      $N$(H)=$2.6\times10^{20}$\,cm$^{-2}$. The model H$_2$ Doppler
      width is $3$\,km\,s$^{-1}$. Also shown is the
      ${\rm{}H}_2+{\rm{}H}$ absorption spectrum (blue) towards $\zeta$
      Oph using the observed column densities for individual $J$
      levels, showing enhanced non-thermal excitation of H$_2$ in the
      higher $J$ levels. \chg{The asterisks
        indicate the $c'(0)$ (Band 20) and $c(0)$ (Band 21) bands,
    respectively, detected
        in absorption towards HD~124314.}}
\label{fig:N2overview}
\end{figure*}
}

\def\placefigCOspectra{ \begin{figure*}[htb] \centering
    \includegraphics[angle=0,width=1\hsize,height=6.5cm]{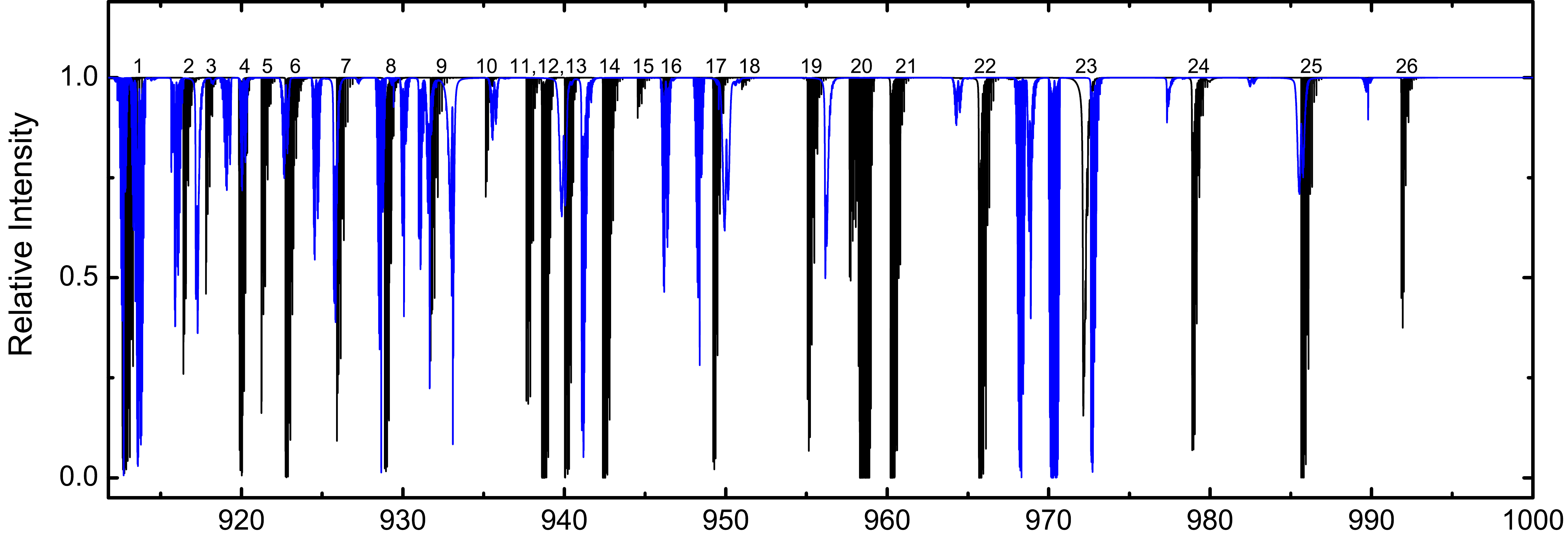}
    \includegraphics[angle=0,width=1\hsize,height=6.5cm]{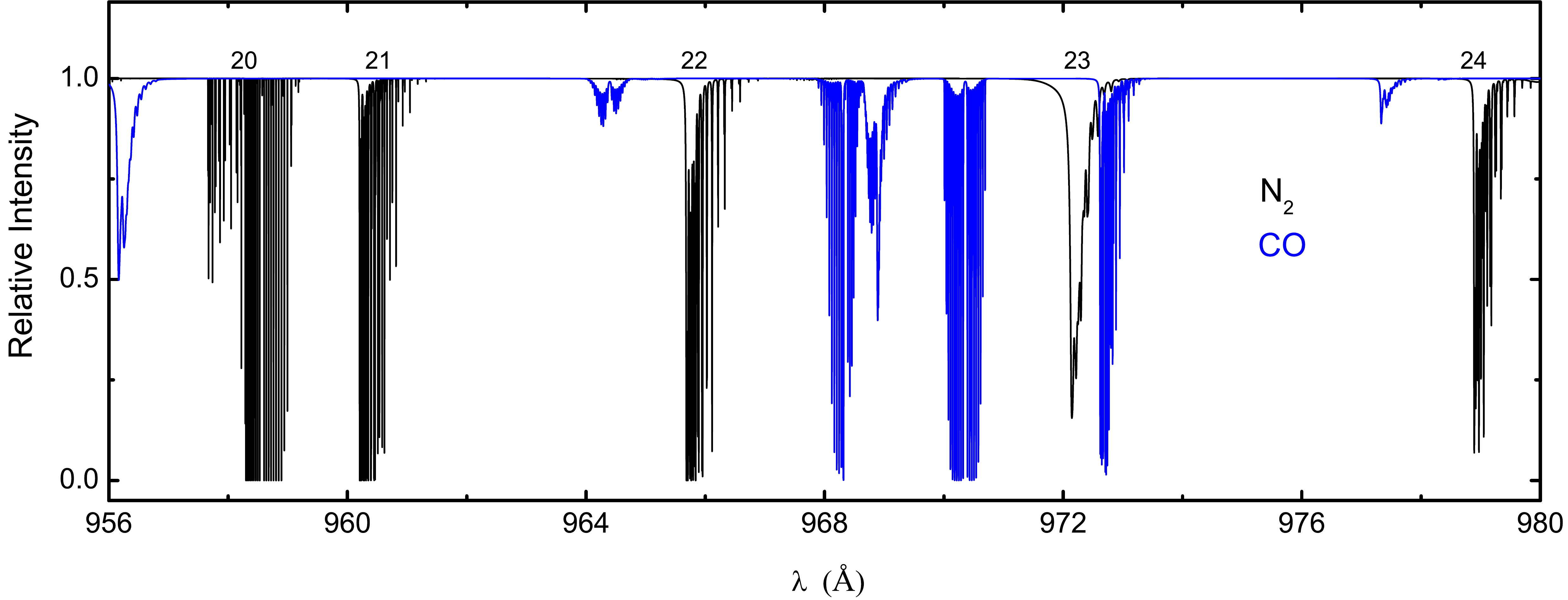}
    \caption{$Top$: Comparison of N$_2$ (black) and CO (blue) model
      absorption spectra between 912 and 1000\,\AA{} assuming an
      excitation temperature of 50\,K for both molecules. The N$_2$
      and CO column densities are both $10^{15}$\,cm$^{-2}$. $Bottom$:
      Blow-up of the above spectra for the wavelength region $956-980$\,\AA{}.}
\label{fig:N2COspectra}
\end{figure*}
}

\def\placefigshielding{
\begin{figure}[htb]
\centering
\includegraphics[angle=0,width=1\hsize]{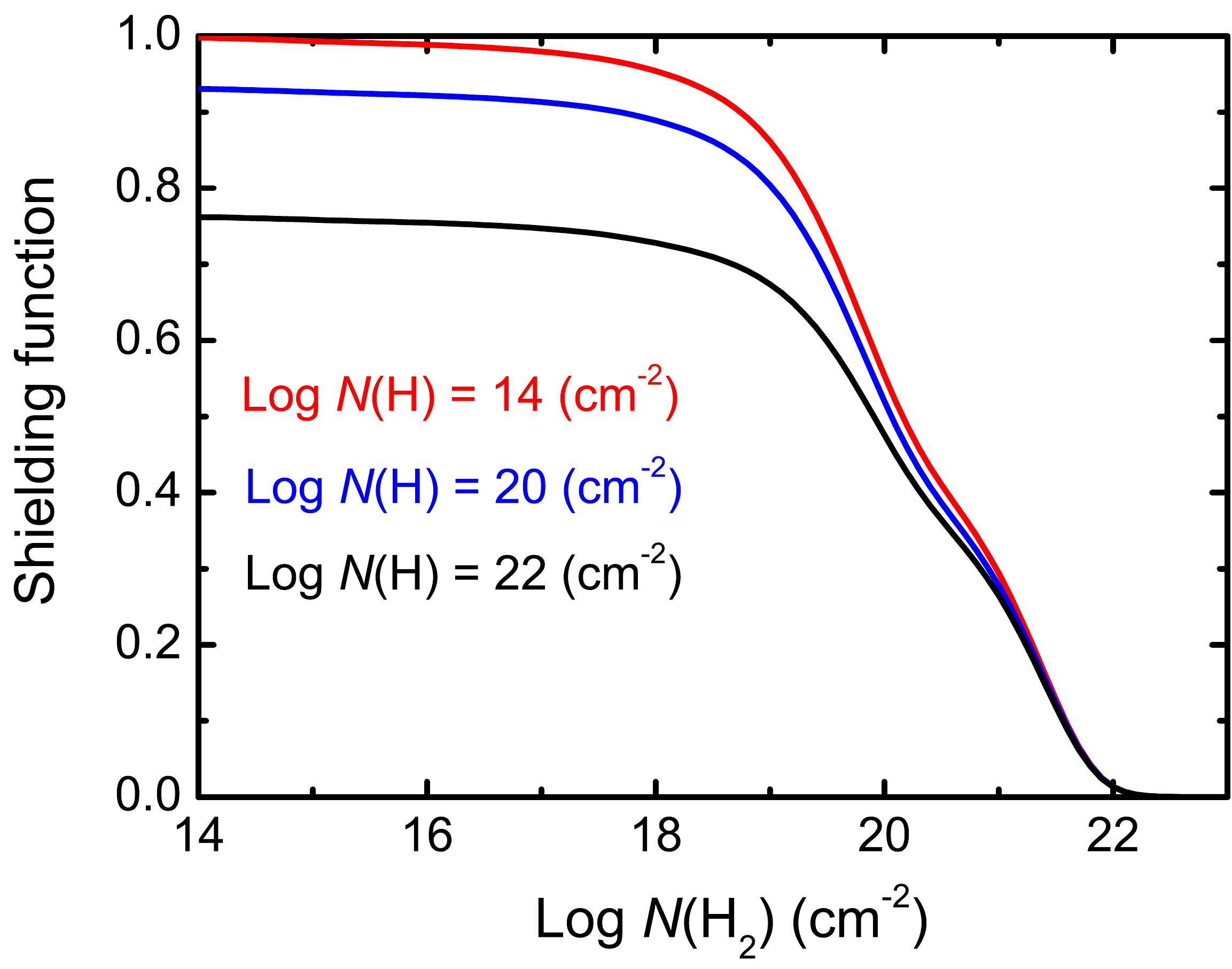}
\caption{Shielding of N$_2$ by ${\rm H}_2+{\rm H}$ as a function of
  H$_2$ column density, $N$(H$_2$), for three different values of
  $N$(H). An excitation temperature of 50\,K is adopted for both N$_2$
  and H$_2$.}
\label{fig:shielding-diff-logN(H)}
\end{figure}
}

\def\placefigshieldingtemperatures{
\begin{figure}[htb]
\centering
\includegraphics[angle=0,width=1\hsize]{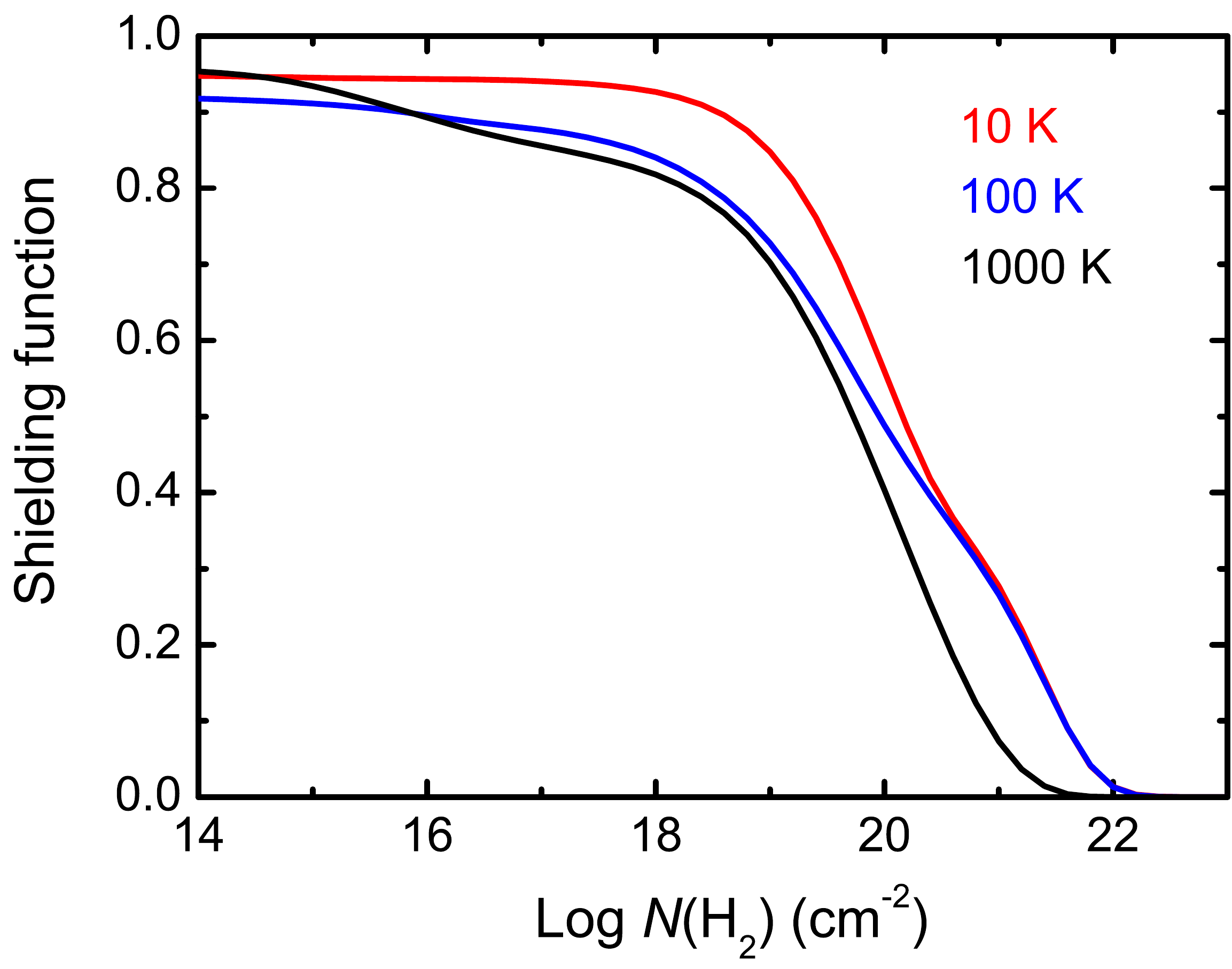}
\caption{Shielding of N$_2$ by ${\rm H}_2+{\rm H}$ as a function of
  H$_2$ column density $N$(H$_2$), for N$_2$ and
  H$_2$ excitation temperatures of 10, 50, 100 and 1000~K.
  The column-density of H is set to $10^{20}$\,cm$^{-2}$ in all cases.}
    \label{fig:shieldingtemperatures}
    \end{figure}
}

\def\placefigselfshielding{
\begin{figure}[htb]
\centering
\includegraphics[angle=0,width=1\hsize]{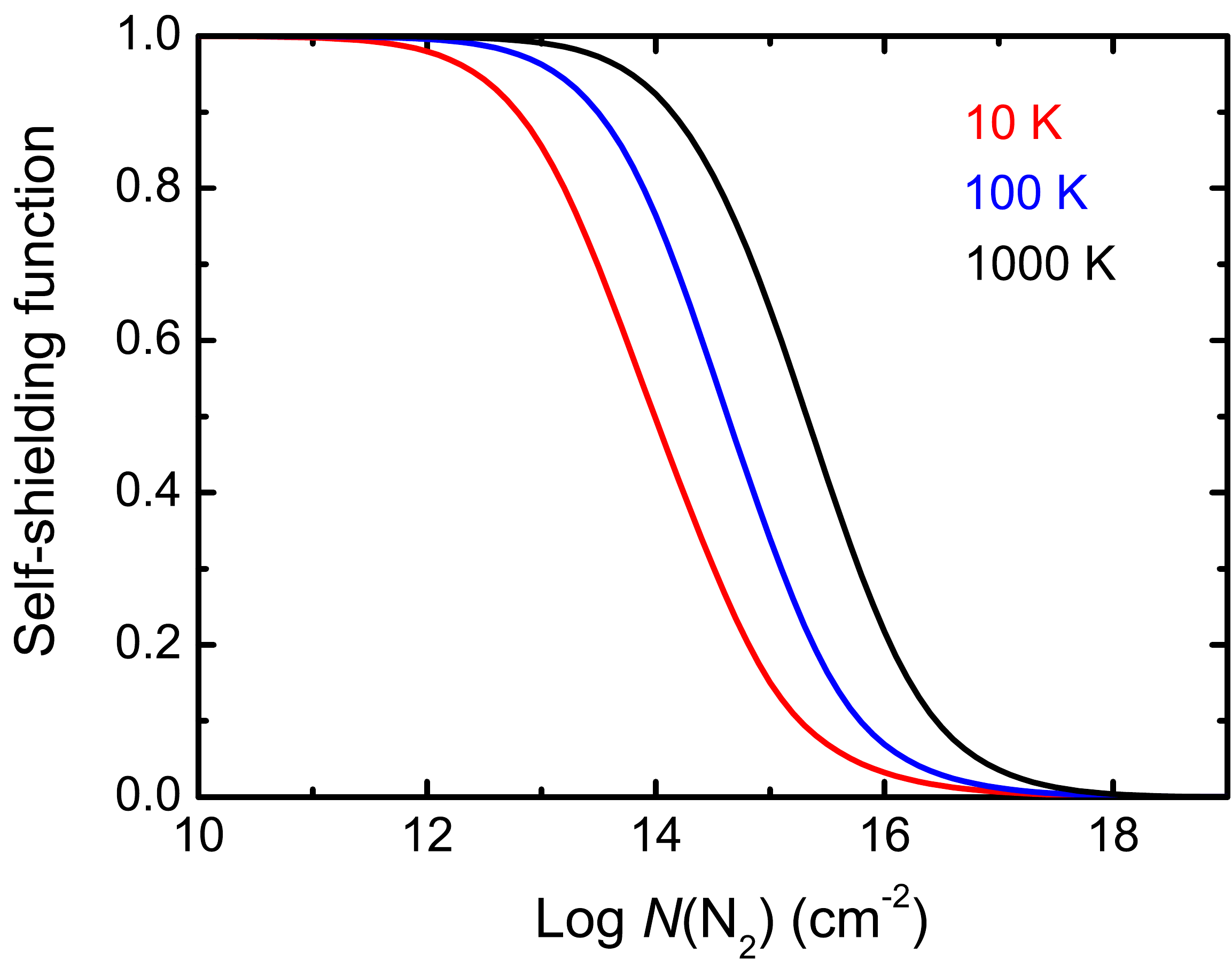}
\caption{N$_2$ self-shielding as a function of column density,
  $N$(N$_2$), for excitation temperatures of 10, 50, 100 and 1000\,K. }
    \label{fig:self-shielding}
    \end{figure}
}

\def\placefigrates{
   \begin{figure}[htb] \centering
    \includegraphics[angle=0,width=1\hsize]{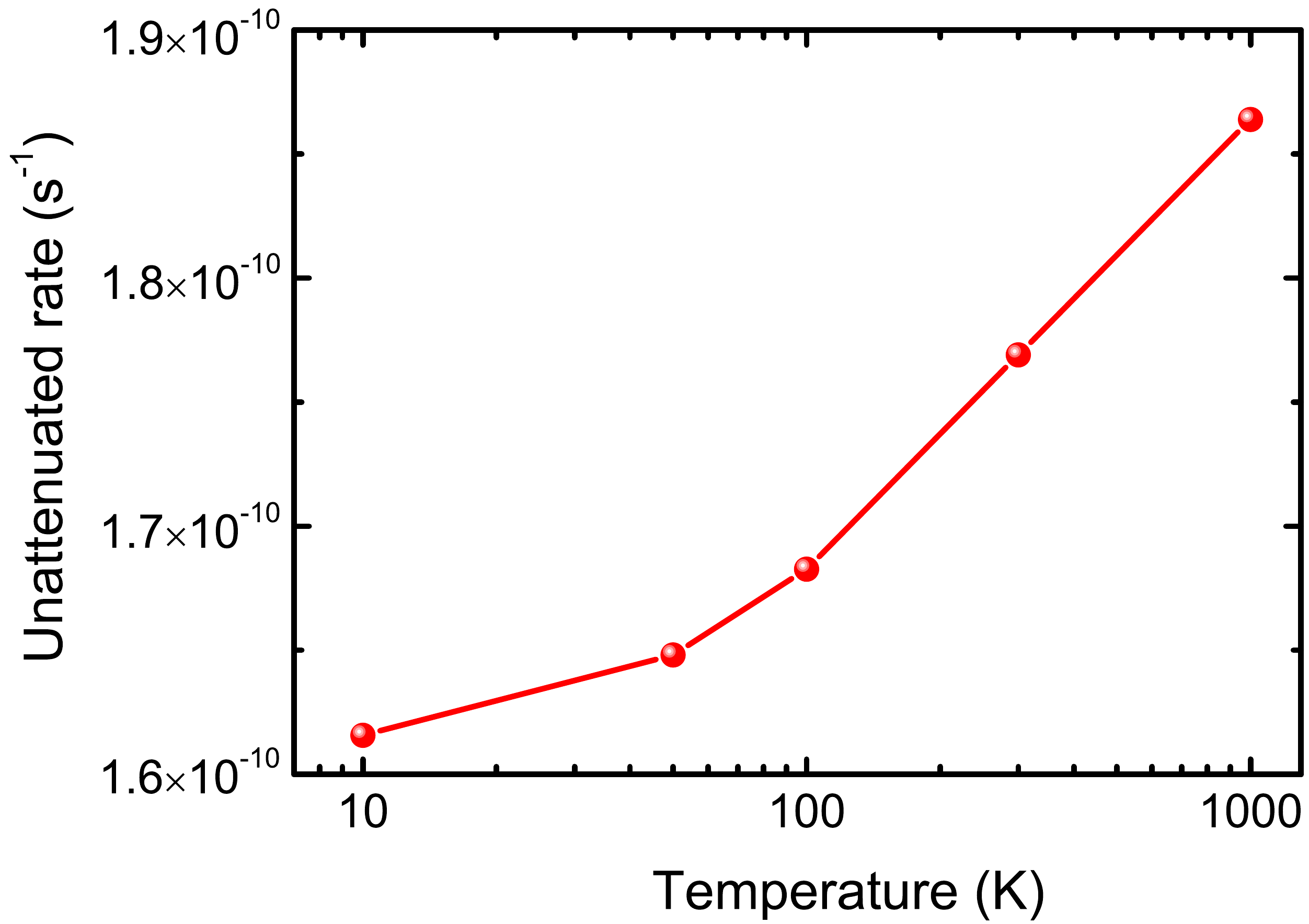}
    \caption{The unattenuated photodissociation rates of N$_2$
      immersed in a Draine (1978) field at various excitation temperatures.}
    \label{fig:rates}
    \end{figure}
}

\def\placefigpartitionfunction{
   \begin{figure}[htb]
\centering
\includegraphics[angle=0,width=1\hsize]{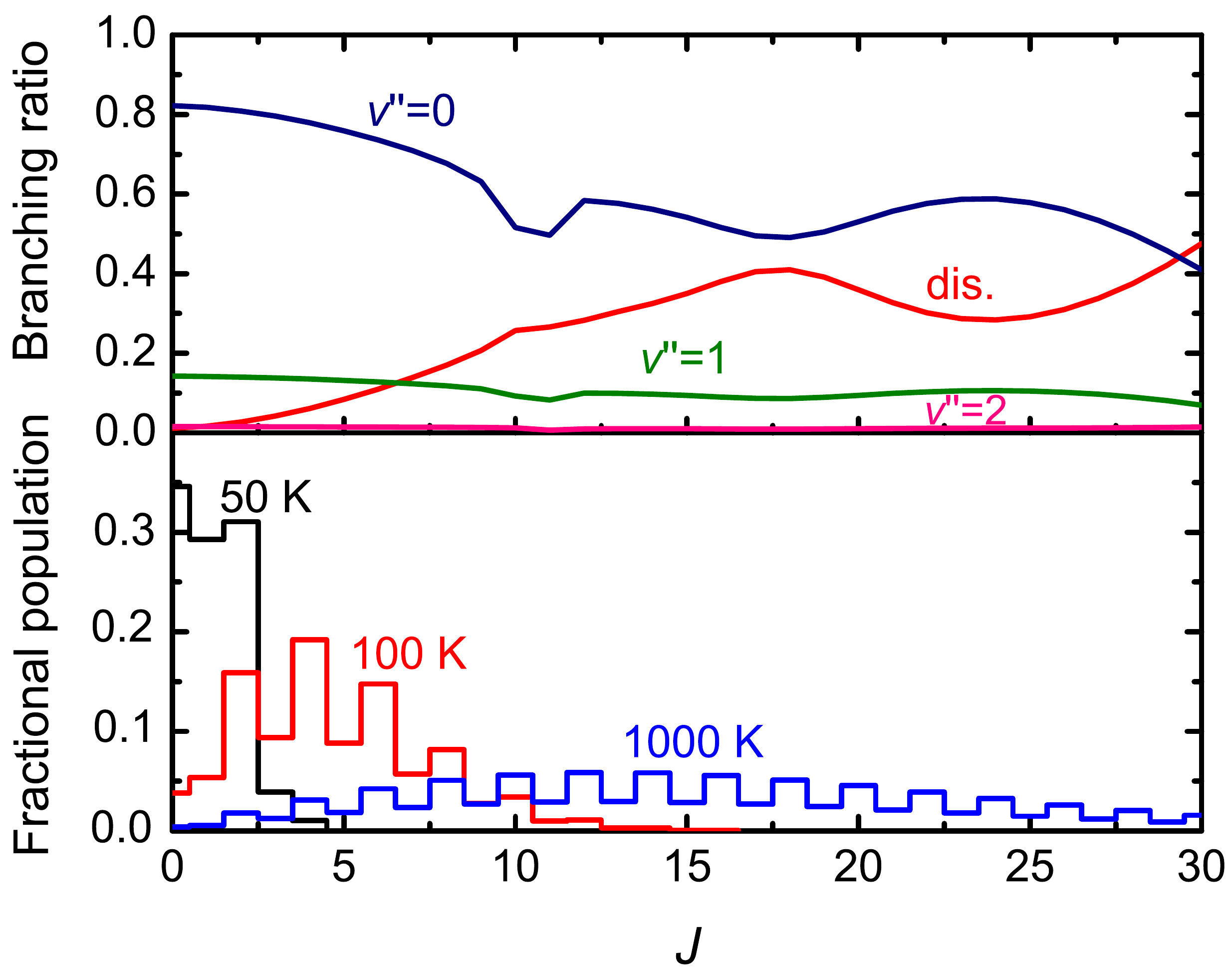}
\caption{\emph{Top:} The branching to various decay channels of the
  $c'(v'=0)$ excited-state of N$_2$ as a function of total rotational
  quantum number, $J$. The figure includes spontaneous emission to
  several non-dissociative ground state vibrational levels ($v''=0,1$
  and 2) and decay due to predissociation (dis.).  \emph{Bottom:}
  Fractional population of the N$_2$ ground state in its lowest
  vibrational level as a function of $J$ and for several excitation
  temperatures. The 2:1 ratio of populations for even:odd $J$ levels
  arises from the combined rotational and nuclear spin statistics.}
    \label{fig:Partition-function}
    \end{figure}
}

\def\placefigcrossections{
     \begin{figure}[htb]
\centering  
\includegraphics[angle=0,width=1\hsize]{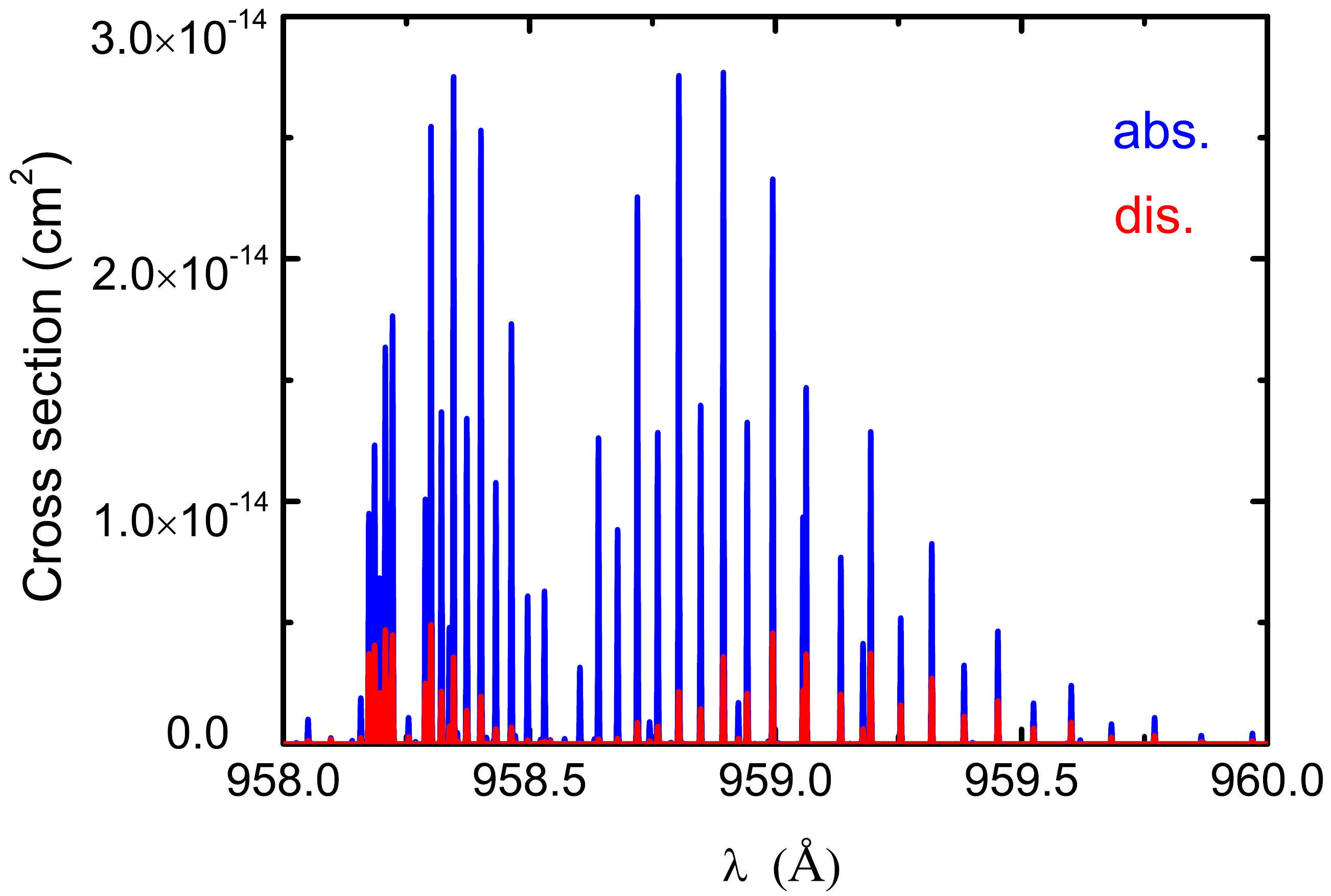}
\caption{The CSE-calculated absorption cross section (blue) of the
  $c'(v'=0)$ level of N$_2$ assuming an excitation temperature of
  300\,K. Also shown is a dissociation cross section (red) which
  has been corrected for the non-unity dissociation efficiency, $\eta_J$, of this
  band (see Fig.~\ref{fig:Partition-function}). }
    \label{fig:cross_section}
    \end{figure}
}

\def\placefigtrans{
     \begin{figure}[htb]
\centering  
\includegraphics[angle=0,width=1\hsize]{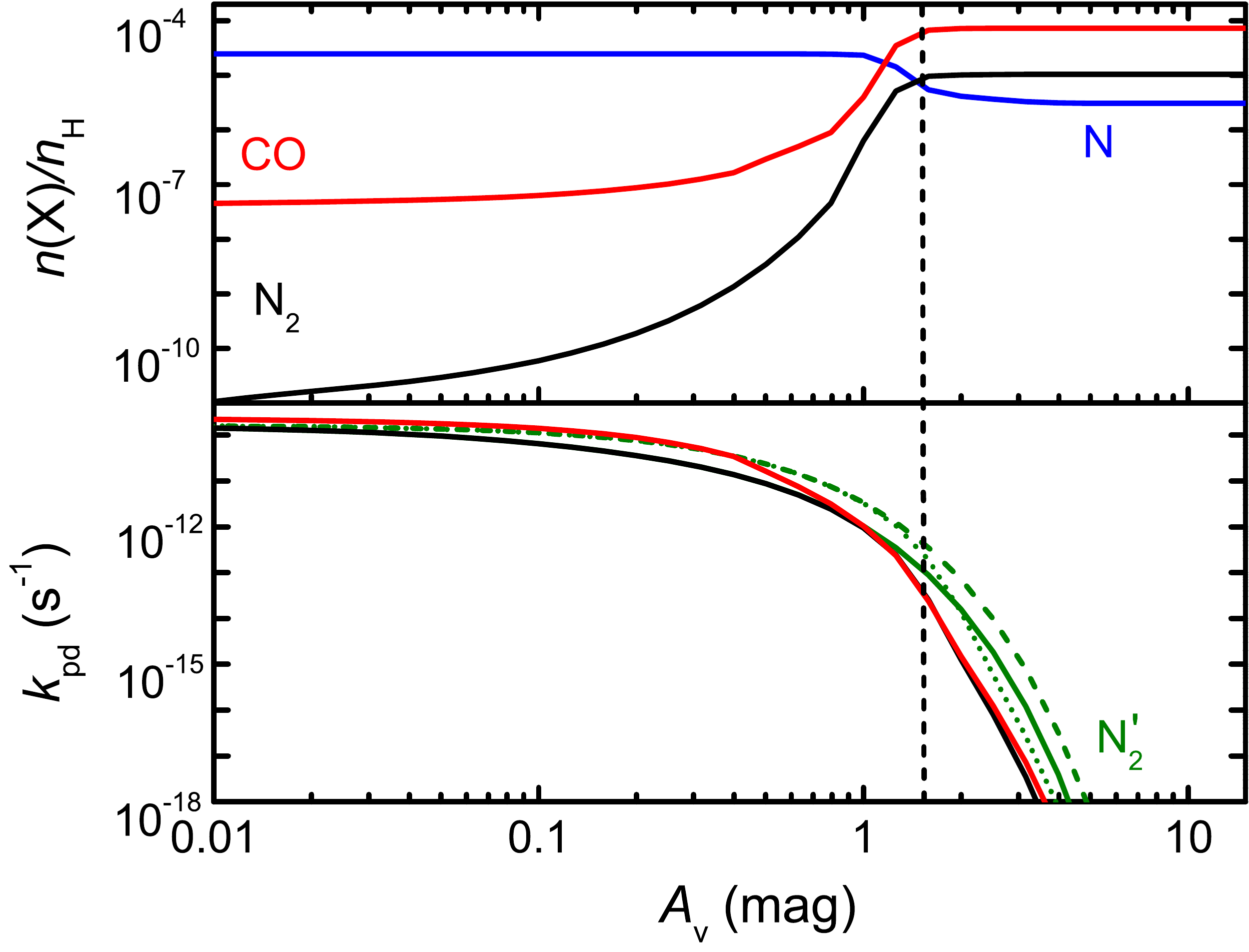}
\caption{$Top$: Relative abundances of N, N$_2$ and CO as a function
  of depth into a translucent cloud with $\chi=1$, $T=30$~K and
  $n_\mathrm{H}=10^3\,\mathrm{cm}^{-3}$. $Bottom$: Photodissociation
  rates of N$_2$ (black) and CO (red) as functions of depth. Alternative
  photodissociation rate curves (${\rm N}^{\stackrel{\prime}{\vspace{1ex}}}_2$, green) consider shielding by dust alone (dashed), dust + self-shielding (dotted), and dust + H + H$_2$ (solid). The conversion from N to N$_2$ occurs at $A_{\rm V}\simeq1.5$\,mag, as shown by the vertical dashed line.
   }
    \label{fig:trans}
    \end{figure}
}

\def\placefigpdr{
      \begin{figure}[htb]
\centering
\includegraphics[angle=0,width=1\hsize]{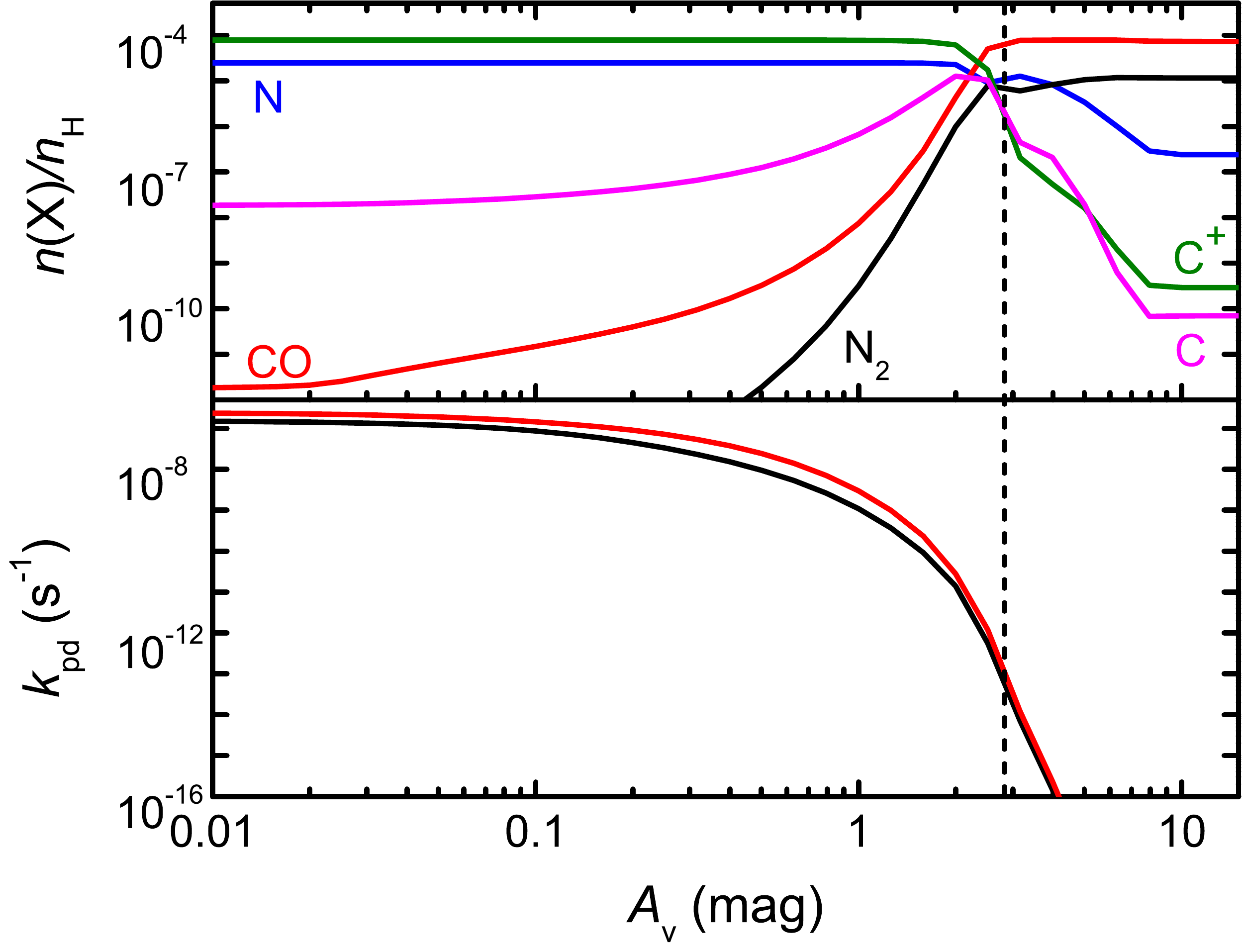}
\caption{$Top$: Relative abundances of N (blue), N$_2$ (black), C
  (magenta), C$^+$ (green) and CO (red) as functions of depth into a
  dense PDR. $Bottom$: Photodissociation rates of N$_2$ and CO as
  functions of depth into a dense PDR. The vertical dashed line
  indicates where He$^+$ and H$_3^+$ take over from photodissociation
  as the main destruction mechanism for N$_2$.}
    \label{fig:pdr}
    \end{figure}
}

\def\placefigdisk{
         \begin{figure}[htb]
\centering
\includegraphics[angle=0,width=1\hsize]{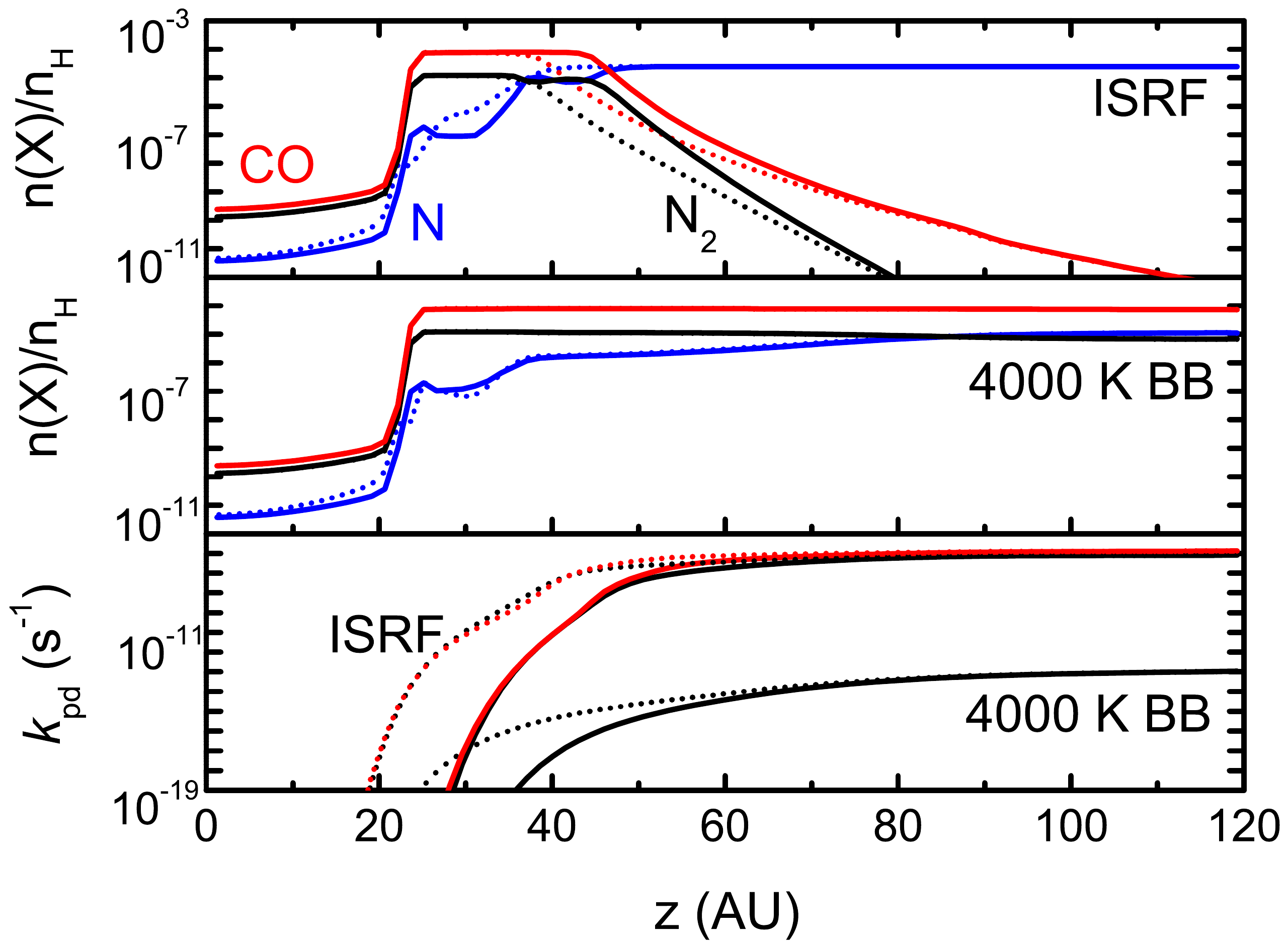}

\caption{$Top$ and $Middle$: Relative abundances of N (blue), N$_2$
  (black) and CO (red) as a function of height in a slice through a
  circumstellar disk exposed to the interstellar radiation field (ISRF)
 scaled by a factor $\chi=$516 and a 4000~K black body field.  $Bottom$
  The relevant photodissociation rates of CO and N$_2$. Solid lines
  are for a grain size of 0.1 $\mu$m, dotted lines for 1 $\mu$m.}
    \label{fig:disk}
    \end{figure}
}

\def\placefigcoldentransgrid{
         \begin{figure}[htb]
\centering
\includegraphics[angle=0,width=1\hsize]{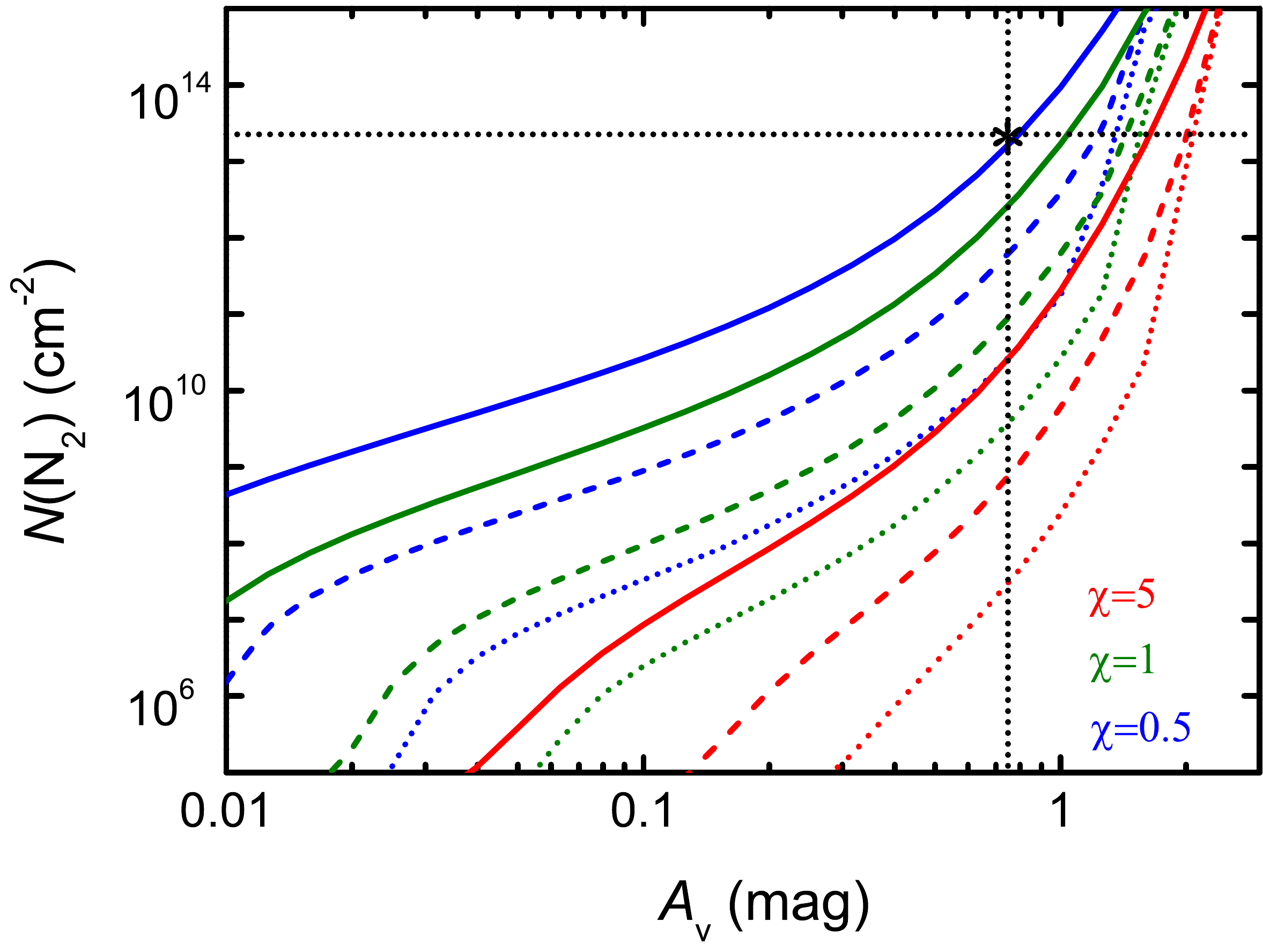}
\caption{Cumulative column density of N$_2$ as a function of
  extinction, $A_{\rm V}$, at an excitation temperature of 30~K. Curves are shown for several different
  models: dotted, dashed, and solid lines indicate $n_\mathrm{H}=100$,
  300 and 1000\,cm$^{-3}$, respectively; blue, green and red lines
  scale the radiation field by $\chi=0.5$, 1 and 5, respectively.  The
  asterisk indicates half the N$_2$ column density and half the extinction
  observed by \citet{Knauth04} in the diffuse cloud toward
  HD~124314.  }
    \label{fig:coldentransgrid}
    \end{figure}
}



\def\placetabcontributions{
\begin{table}
\caption{Contributions of different bands to N$_2$ photodissociation at 50~K at the edge (unattenuated photodissociation) and in the center of the $\zeta$ Oph diffuse cloud.}
\label{tab:contribution}
\centering
\begin{tabular}{cccccc}
\hline\hline
     & Excited &                 & Edge & Center &            \\
Band & state\tablefootmark{a}   & $\lambda$ (\AA) & (\%) & (\%)   & Shielding  \\
\hline
1&	$o(2)$	&	911.7-915.1	&	2.41	&	0.07	&	0.01	\\
2&	$b(11)$	&	915.1-917.4	&	0.58	&	0.47	&	0.34	\\
3&	$b'(7)$	&	917.4-919.0	&	0.09	&	0.06	&	0.30	\\
4&	$c(2)$	&	919.0-920.8	&	1.63	&	3.65	&	0.95	\\
5&	$c'(2)$	&	920.8-922.4	&	0.22	&	0.51	&	0.97	\\
6&	$b(10)$	&	922.4-924.7	&	1.74	&	3.73	&	0.91	\\
7&	$b'(6)$	&	924.7-927.7	&	0.33	&	0.66	&	0.84	\\
8&	$o(1)+b(9)$& 927.7-930.8 &	3.65	&	2.31	&	0.27	\\
9&	$b'(5)$	&	930.8-933.9	&	0.23	&	0.10	&	0.18	\\
10&	$b(8)$	&	933.9-936.6	&	0.10	&	0.22	&	0.96	\\
11&	$b'(4)$	&	936.6-938.5	&	0.47	&	0.52	&	0.47	\\
12&	$c(1)$	&	938.5-939.6	&	9.31	&	5.28	&	0.24	\\
13&	$c'(1)$	&	939.6-941.5	&	1.67	&	3.47	&	0.88	\\
14&	$b(7)$	&	941.5-944.1	&	5.01	&	11.04	& 0.93	\\
15&	$b'(3)$	&	944.1-945.7	&	0.03	&	0.05	&	0.77	\\
16&	$o(0)$	&	945.7-947.9	&	0.05 &	0.00	&	0.00   \\
17&	$b(6)$	&	947.9-950.6	&	1.23	&	2.00	&	0.69	\\
18&	$b'(2)$	&	950.6-953.1	&	0.01	&	0.02	&	0.93	\\
19&	$b(5)$	&	953.1-956.8	&	1.04	&	0.41	&	0.17	\\
20&	$c'(0)+b'(1)$&	956.8-959.5	&	2.79	&	6.21	&	0.94	\\
21&	$c(0)$	&	959.5-962.9	&	17.00	&	36.80	&	0.92	\\
22&	$b(4)+b'(0)$&	962.9-969.4	&	22.51	&	2.02	&	0.04	\\
23&	$b(3)$	&	969.4-976.2	&	16.05	&	0.51	&	0.01	\\
24&	$b(2)$	&	976.2-983.1	&	9.28	&	19.83	&	0.91	\\
25&	$b(1)$	&	983.1-988.9	&	1.34	&	0.00	&	0.00	\\
26&	$b(0)$	&	988.9-1000.0 &	1.24	&	0.05	&	0.02	\\
\hline
\end{tabular}
\tablefoot{\tablefoottext{a}{\chg{The $b$, $c$, and $o$ levels have ${}^1\Pi_u$ symmetry; $b'$ and $c'_4$ have  ${}^1\Sigma_u^+$ symmetry.}}}
\end{table}
}

\def\placetabshielding{
\begin{table}
\caption{Two-dimensional shielding functions
  $\Theta$[$N$(H),$N$(H$_2$)] assuming an excitation temperature of 50\,K. The notation $x(-y)$ indicates $x\times 10^{-y}$.} 
\label{tab:shielding-logN(H)=14,20,22}
\centering
\begin{tabular}{cccc}
\hline\hline
 log $N$(H$_2$)  &            & log $N$(H) (cm$^{-2}$)  &            \\
\cline{2-4}
     (cm$^{-2}$) & 14         & 20          & 22          \\
\hline
14.0 & 9.979(-1) & 9.307(-1) & 7.621(-1) \\
14.2 & 9.972(-1) & 9.301(-1) & 7.617(-1) \\
14.4 & 9.964(-1) & 9.293(-1) & 7.610(-1) \\
14.6 & 9.955(-1) & 9.284(-1) & 7.603(-1) \\
14.8 & 9.945(-1) & 9.274(-1) & 7.595(-1) \\
15.0 & 9.934(-1) & 9.264(-1) & 7.586(-1) \\
15.2 & 9.923(-1) & 9.254(-1) & 7.577(-1) \\
15.4 & 9.913(-1) & 9.245(-1) & 7.569(-1) \\
15.6 & 9.904(-1) & 9.237(-1) & 7.562(-1) \\
15.8 & 9.896(-1) & 9.228(-1) & 7.555(-1) \\
16.0 & 9.885(-1) & 9.219(-1) & 7.546(-1) \\
16.2 & 9.873(-1) & 9.207(-1) & 7.536(-1) \\
16.4 & 9.859(-1) & 9.193(-1) & 7.524(-1) \\
16.6 & 9.841(-1) & 9.177(-1) & 7.509(-1) \\
16.8 & 9.821(-1) & 9.157(-1) & 7.492(-1) \\
17.0 & 9.797(-1) & 9.134(-1) & 7.472(-1) \\
17.2 & 9.767(-1) & 9.105(-1) & 7.447(-1) \\
17.4 & 9.729(-1) & 9.069(-1) & 7.418(-1) \\
17.6 & 9.681(-1) & 9.023(-1) & 7.380(-1) \\
17.8 & 9.619(-1) & 8.964(-1) & 7.335(-1) \\
18.0 & 9.541(-1) & 8.890(-1) & 7.280(-1) \\
18.2 & 9.443(-1) & 8.800(-1) & 7.217(-1) \\
18.4 & 9.319(-1) & 8.685(-1) & 7.141(-1) \\
18.6 & 9.154(-1) & 8.534(-1) & 7.046(-1) \\
18.8 & 8.927(-1) & 8.326(-1) & 6.916(-1) \\
19.0 & 8.615(-1) & 8.040(-1) & 6.738(-1) \\
19.2 & 8.198(-1) & 7.658(-1) & 6.496(-1) \\
19.4 & 7.663(-1) & 7.164(-1) & 6.175(-1) \\
19.6 & 7.012(-1) & 6.563(-1) & 5.766(-1) \\
19.8 & 6.281(-1) & 5.887(-1) & 5.280(-1) \\
20.0 & 5.540(-1) & 5.201(-1) & 4.758(-1) \\
20.2 & 4.875(-1) & 4.583(-1) & 4.259(-1) \\
20.4 & 4.333(-1) & 4.078(-1) & 3.830(-1) \\
20.6 & 3.886(-1) & 3.660(-1) & 3.457(-1) \\
20.8 & 3.446(-1) & 3.251(-1) & 3.081(-1) \\
21.0 & 2.940(-1) & 2.781(-1) & 2.645(-1) \\
21.2 & 2.327(-1) & 2.213(-1) & 2.113(-1) \\
21.4 & 1.630(-1) & 1.561(-1) & 1.497(-1) \\
21.6 & 9.495(-2) & 9.178(-2) & 8.843(-2) \\
21.8 & 4.260(-2) & 4.164(-2) & 4.030(-2) \\
22.0 & 1.375(-2) & 1.359(-2) & 1.320(-2) \\
22.2 & 3.210(-3) & 3.200(-3) & 3.120(-3) \\
22.4 & 5.737(-4) & 5.734(-4) & 5.600(-4) \\
22.6 & 6.110(-5) & 6.108(-5) & 5.969(-5) \\
22.8 & 2.146(-6) & 2.146(-6) & 2.094(-6) \\
23.0 & 1.151(-8) & 1.151(-8) & 1.118(-8) \\
\hline
\end{tabular}
\end{table}
}

\def\placetabBB{
\begin{table}
  \caption{The unattenuated photodissociation rates of N$_2$ (excitation temperature 50\,K) in a blackbody radiation field at various temperatures, $T_{\rm BB}$. }
\label{tab:BB-kpd0}
\centering
\begin{tabular}{ccc}
\hline\hline
$T_{\rm BB}$(\,K)\tablefootmark & $k^0_{\rm pd}$\tablefootmark{a}  (s$^{-1}$) & Previous \\ 
\hline
4000 & 2.18(-16) & 3.0(-16)\tablefootmark{b} \\
6000 & 9.96(-14) & -            \\
8000 & 1.90(-12) & -            \\
10\,000 & 1.03(-11) & 1.4(-11)\tablefootmark{b} \\
20\,000 & 1.97(-10) & -           \\                [1ex] 
Draine & 1.65(-10) & 2.3(-10)\tablefootmark{c} \\
\hline
\end{tabular}
\tablefoot{
\tablefoottext{a}{All radiation fields have been normalised to a \citet{Draine78} field over the interval 912--2050\,\AA.} \ \
\tablefoottext{b}{\citet{vanDishoeck06}}\ \ \
\tablefoottext{c}{\citet{vanDishoeck88}} \\ \
}
\end{table}
}

\def\placetabcodust{
\begin{table*}
  \caption{ Comparison of the shielding of $^{14}$N$_2$ and $^{12}$CO
    by ${\rm{}H}_2+{\rm{}H}$ and dust for a range of extinction,
    $A_{\rm V}$, at 10~K and
    taking $N({\rm H})=5\times 10^{20}\,{\rm{}cm}^{-2}$.\tablefootmark{a} }
\label{tab:shielding-co-N2-dust}
\centering
\begin{tabular}{cccccccc}
\hline\hline

\multicolumn{4}{c}{Shielding due to ${\rm H}_2+{\rm H}$}     &
\hspace{7ex} &
\multicolumn{2}{c}{Shielding due to dust}  \\

$A_{\rm V}$ & $\log N({\rm H}_2)$ (cm$^{-2}$)    & $^{14}$N$_2$ & $^{12}$CO\tablefootmark{b}  &    & Interstellar  & Protoplanetary  \\
\hline
0.31   & 0              & 8.916(-1)    & 0.9                      &   & 3.318(-1)         & 8.290(-1)            \\
0.32   & 19             & 8.098(-1)    & 8.176(-1)                  &   & 3.175(-1)         & 8.228(-1)            \\
0.44   & 20             & 5.543(-1)    & 7.223(-1)                  &   & 2.134(-1)         & 7.691(-1)            \\
1.56   & 21             & 2.759(-1)    & 3.260(-1)                  &   & 4.023(-3)         & 3.916(-1)            \\
12.8  & 22             & 1.346(-2)    & 1.108(-2)                  &   & 0                 & 4.585(-4)            \\
125.3 & 23             & 2.343(-8)    & 3.938(-7)                  &   & 0                 & 0                    \\
\hline

\end{tabular}
\tablefoot{
  \tablefoottext{a}The unattenuated
    photodissociation rates of $^{14}$N$_2$ and $^{12}$CO are $1.65
    \times 10^{-10}$ and $2.59 \times 10^{-10}$ s$^{-1}$,
    respectively.} \tablefoottext{b}{The CO values are from \citet{Visser09}.}
\end{table*}
}

%
%
\def\placetabselfshielding{
\begin{table}
  \caption{N$_2$ self-shielding as a function of column density,
  $N$(N$_2$), for excitation temperatures of 10, 100 and 1000\,K. The notation $x(-y)$ indicates $x\times 10^{-y}$.}
\label{tab:online-selfshielding}
\centering
\begin{tabular}{cccc}
\hline\hline
 log $N$(N$_2$)  &            & Temperature (K)  &            \\
\cline{2-4}
     (cm$^{-2}$) & 10         & 100          & 1000          \\
\hline
10.0 & 9.998(-1) & 1.000(0)  & 1.000(0)  \\
10.2 & 9.997(-1) & 9.999(-1) & 1.000(0)  \\
10.4 & 9.995(-1) & 9.999(-1) & 1.000(0)  \\
10.6 & 9.991(-1) & 9.998(-1) & 1.000(0)  \\
10.8 & 9.986(-1) & 9.997(-1) & 9.999(-1) \\
11.0 & 9.979(-1) & 9.996(-1) & 9.999(-1) \\
11.2 & 9.966(-1) & 9.994(-1) & 9.999(-1) \\
11.4 & 9.947(-1) & 9.990(-1) & 9.998(-1) \\
11.6 & 9.917(-1) & 9.984(-1) & 9.996(-1) \\
11.8 & 9.870(-1) & 9.974(-1) & 9.994(-1) \\
12.0 & 9.798(-1) & 9.960(-1) & 9.991(-1) \\
12.2 & 9.691(-1) & 9.936(-1) & 9.986(-1) \\
12.4 & 9.533(-1) & 9.900(-1) & 9.977(-1) \\
12.6 & 9.305(-1) & 9.844(-1) & 9.964(-1) \\
12.8 & 8.985(-1) & 9.757(-1) & 9.943(-1) \\
13.0 & 8.554(-1) & 9.627(-1) & 9.910(-1) \\
13.2 & 8.001(-1) & 9.436(-1) & 9.859(-1) \\
13.4 & 7.333(-1) & 9.163(-1) & 9.781(-1) \\
13.6 & 6.579(-1) & 8.785(-1) & 9.662(-1) \\
13.8 & 5.778(-1) & 8.282(-1) & 9.486(-1) \\
14.0 & 4.967(-1) & 7.643(-1) & 9.235(-1) \\
14.2 & 4.169(-1) & 6.877(-1) & 8.891(-1) \\
14.4 & 3.400(-1) & 6.022(-1) & 8.440(-1) \\
14.6 & 2.680(-1) & 5.130(-1) & 7.875(-1) \\
14.8 & 2.037(-1) & 4.244(-1) & 7.197(-1) \\
15.0 & 1.505(-1) & 3.394(-1) & 6.412(-1) \\
15.2 & 1.100(-1) & 2.609(-1) & 5.546(-1) \\
15.4 & 8.073(-2) & 1.928(-1) & 4.641(-1) \\
15.6 & 5.945(-2) & 1.384(-1) & 3.747(-1) \\
15.8 & 4.374(-2) & 9.802(-2) & 2.911(-1) \\
16.0 & 3.214(-2) & 6.920(-2) & 2.175(-1) \\
16.2 & 2.361(-2) & 4.886(-2) & 1.567(-1) \\
16.4 & 1.731(-2) & 3.454(-2) & 1.100(-1) \\
16.6 & 1.267(-2) & 2.441(-2) & 7.613(-2) \\
16.8 & 9.250(-3) & 1.717(-2) & 5.229(-2) \\
17.0 & 6.720(-3) & 1.199(-2) & 3.557(-2) \\
17.2 & 4.870(-3) & 8.280(-3) & 2.387(-2) \\
17.4 & 3.510(-3) & 5.640(-3) & 1.574(-2) \\
17.6 & 2.520(-3) & 3.810(-3) & 1.013(-2) \\
17.8 & 1.790(-3) & 2.570(-3) & 6.340(-3) \\
18.0 & 1.270(-3) & 1.740(-3) & 3.850(-3) \\
18.2 & 8.906(-4) & 1.190(-3) & 2.280(-3) \\
18.4 & 6.247(-4) & 8.066(-4) & 1.310(-3) \\
18.6 & 4.359(-4) & 5.425(-4) & 7.183(-4) \\
18.8 & 2.998(-4) & 3.613(-4) & 3.737(-4) \\
19.0 & 2.031(-4) & 2.386(-4) & 1.820(-4) \\
\hline
\end{tabular}
\end{table}
}



\abstract
{Molecular nitrogen is one of the key species in the
    chemistry of interstellar clouds and protoplanetary disks, but its
    photodissociation under interstellar conditions has never been
    properly studied. The partitioning of nitrogen between N and N$_2$
    controls the formation of more complex prebiotic
    nitrogen-containing species.}
  {The aim of this work is to gain a better understanding of the
  interstellar N$_2$ photodissociation processes based on recent
  detailed theoretical and experimental work and to provide accurate
  rates for use in chemical models. }
{We used an approach similar to that adopted for CO in which we
  simulated the full high-resolution line-by-line absorption +
  dissociation spectrum of N$_2$ over the relevant 912--1000 \AA\
  wavelength range, by using a quantum-mechanical model which solves
  the coupled-channels Schr\"odinger equation. The simulated N$_2$ spectra were
  compared with the absorption spectra of H$_2$, H, CO, and dust to
  compute photodissociation rates in various radiation fields and
  shielding functions. The effects of the new rates in interstellar
  cloud models were illustrated for diffuse and translucent clouds, a
  dense photon dominated region and a protoplanetary disk.}
{The unattenuated photodissociation rate in the Draine (1978, ApJS, 36, 595)
  radiation field assuming an N$_2$ excitation temperature of 50\,K is
  $1.65\times 10^{-10}$ s$^{-1}$, with an uncertainty of only 10\%.
  Most of the photodissociation occurs through bands in the 957--980
  \AA\ range. The N$_2$ rate depends slightly on the temperature
  through the variation of predissociation probabilities with
  rotational quantum number for some bands. Shielding functions are
  provided for a range of H$_2$ and H column densities, with H$_2$
  being much more effective than H in reducing the N$_2$ rate inside a
  cloud. Shielding by CO is not effective.  The new rates are 28\%
  lower than the previously recommended values. Nevertheless, diffuse
  cloud models still fail to reproduce the possible detection of
  interstellar N$_2$ except for unusually high densities and/or low
  incident UV radiation fields.  The transition of N $\to$ N$_2$
  occurs at nearly the same depth into a cloud as that of
  $\mathrm{C}^+\to\mathrm{C}\to\mathrm{CO}$.  The orders-of-magnitude
  lower N$_2$ photodissociation rates in clouds exposed to black-body
  radiation fields of only 4000 K can qualitatively explain the lack
  of active nitrogen chemistry observed in the inner disks around cool
  stars.}
  {Accurate photodissociation rates for N$_2$ as a function of depth
  into a cloud are now available that can be applied to a wide variety
  of astrophysical environments.}


\maketitle

%
%



\section{Introduction} \label{sec:intro}

Nitrogen is one of the most abundant elements in the universe and an
essential ingredient for building prebiotic organic molecules. In
interstellar clouds, its main gas-phase reservoirs are N and N$_2$,
with the balance between these species determined by the balance of
the chemical reactions that form and destroy N$_2$. If nitrogen is
primarily in atomic form, a rich nitrogen chemistry can occur leading
to ammonia, nitriles and other nitrogen compounds. On the other hand,
little such chemistry ensues if nitrogen is locked up in the very
stable N$_2$ molecule. The latter situation is similar to that of
carbon with few carbon-chain molecules being produced when most of the
volatile carbon is locked up in CO \citep{Langer89,Bettens95}.

Direct observation of extrasolar N$_2$ is difficult because, unlike
CO, it lacks strong pure rotational or vibrational lines. N$_2$ is well
studied at various locations within our solar system through its
electronic transitions at ultraviolet wavelengths
\citep[e.g.,][]{Strobel82,Meier91,Wayne00,Liang07} and a detection in
interstellar space has been claimed through UV absorption lines in a
diffuse cloud toward the bright background star \chg{HD~124314} \citep{Knauth04}. In
dense clouds well shielded from UV radiation, most nitrogen is
expected to exist as N$_2$ \citep[e.g.,][]{Herbst73,Woodall07} but can
only be detected indirectly through the protonated ion N$_2$H$^+$
\citep{Turner74,Herbst77} or its deuterated form
N$_2$D$^+$. N$_2$H$^+$ emission is indeed widely observed in dense
cores \citep[e.g.,][]{Bergin02,Crapsi05}, star-forming regions
\citep[][]{Fontani11,Tobin12}, protoplanetary disks
\citep[][]{Dutrey07,Oberg10} and external galaxies
\citep[][]{Mauersberger91,Meier05,Muller11}.

Photodissociation is the primary destruction route of N$_2$ in any
region where UV photons are present. Current models of diffuse and
translucent interstellar clouds are unable to reproduce the possible
detection of N$_2$ for one such cloud \citep{Knauth04}.  One possible
explanation is that the adopted N$_2$ photodissociation rate is
incorrect.  Even in dense cores, not all nitrogen appears to have been
transformed to molecular form \citep{Maret06,Daranlot12}. Observations of HCN in
the surface layers of protoplanetary disks suggest that the nitrogen
chemistry is strongly affected by whether or not a star has sufficiently hard
UV radiation to photodissociate N$_2$ \citep{Pascucci09}.  Thus, not
only the absolute photodissociation rate but also its wavelength
dependence is relevant.  All of these astronomical puzzles make a
thorough study of the interstellar N$_2$ photodissociation very timely.

In contrast with many other simple diatomic molecules, the
photodissociation of interstellar N$_2$ has never been properly
studied \citep{vanDishoeck88,vanDishoeck06}. The reason for this is
that the photodissociation of N$_2$, similarly to CO, is initiated by
line absorptions at wavelengths below 1000\,\AA{} (1100 \AA{}
for CO), where high-resolution laboratory spectroscopy has been
difficult. To compute the absolute rate and to treat the depth
dependence of the photodissociation correctly, the full
high-resolution spectrum of the dissociating transitions needs to be
known. Because the absorbing lines become optically thick for modest
N$_2$ column densities, the molecule can shield itself against the
dissociating radiation deeper into the cloud. Moreover, these lines
can be shielded by lines of more abundant species such as H, H$_2$ and
CO. Until recently, accurate N$_2$ molecular data to simulate these
processes were not available. Thanks to a concerted laboratory
\citep[e.g.,][]{ajello_etal1989,helm_etal1993,sprengers_etal2003, sprengers_etal2004b,
  sprengers_etal2005b, Stark08, lewis_etal2008a, heays_etal2009,
  Heays11} and theoretical
\citep[e.g.,][]{spelsberg_meyer2001, Lewis05b, Lewis05a,
  haverd_etal2005, lewis_etal2008c, lewis_etal2008b, ndome_etal2008}
effort over the last two decades, this information is now available.

In this paper, we use a high resolution model spectrum of the absorption and
dissociation of N$_2$ together with simulated spectra of H, H$_2$ and
CO to determine the interstellar N$_2$ photodissociation rate and its
variation with depth into a cloud.
The effect of the new rates on
interstellar N$_2$ abundances is illustrated through a few
representative cloud models. In particular, the N$_2$ abundance in
diffuse and translucent clouds is revisited to investigate whether the
new rates alleviate the discrepancy between models and the possible
detection of N$_2$ in one cloud \citep{Knauth04}. The data presented
here can be applied to a wide range of astrochemical models, including
interstellar clouds in the local and high redshift universe,
protoplanetary disks and exo-planetary atmospheres.  The
$^{14}$N$^{15}$N photodissociation rate and isotope selective
interstellar processes will be discussed in an upcoming paper (Heays
et al. in prep.) and have been discussed in the context of the
chemistry of Titan by \citet{Liang07}.

\section{Photodissociation processes of N$_2$} \label{sec:photo}
\subsection{Photoabsorption and photodissociation spectrum}

The closed-shell diatomic molecule N$_2$ has a dissociation energy of
78\,715~cm$^{-1}$ (9.76 eV, 1270\ \AA) \citep{huber_herzberg1979},
making it one of the most stable molecules in
nature. Electric-dipole-allowed photoabsorption and predissociation in
N$_2$ starts only in the extreme ultraviolet spectral region, at
wavelengths shorter than 1000\ \AA.  The molecular-orbital (MO)
configuration of the $X\,{^1\Sigma_g^+}$ ground state of N$_2$ is
\begin{equation}
  (1\sigma_g)^2(1\sigma_u)^2(2\sigma_g)^2(2\sigma_u)^2(1\pi_u)^4(3\sigma_g)^2.
\end{equation}

\placefigpotentials

Electric-dipole-allowed transitions from the ground state access only
states of $^1\Pi_u$ and $^1\Sigma_u^+$ symmetry. In the region below
the cutoff energy of the interstellar radiation field of
110\,000~cm$^{-1}$ (13.6 eV, 912\ \AA), five such states are
accessible: the $c^{\prime}$ and $b^{\prime}\,{^1\Sigma_u^+}$ states,
and the $c$, $o$, and $b\,{^1\Pi_u}$ states.  The $c^{\prime}$, $c$,
and $o$ states \chg{(sometimes labelled $c'_4$, $c_3$, and $o_3$;
  respectively)} have Rydberg character, the relevant transitions
corresponding to single-electron excitations from the $3\sigma_g$ or
$1\pi_u$ orbitals into a Rydberg orbital. On the other hand, the
$b^{\prime}$ and $b$ states are valence states of mixed MO
configurations accessed by transitions in which one or two electrons
are excited into antibonding orbitals. The relevant potential-energy
curves (PECs) for these $^1\Pi_u$ and $^1\Sigma_u^+$ states are shown
in Fig.~\ref{fig:N2potentials}, in blue and black, respectively.  The
$c^{\prime}$ and $c$ states, whose PECs have the smallest equilibrium
internuclear distance in Fig.~\ref{fig:N2potentials}, are the first
members of Rydberg series converging on the ground state of the
N$_2^+$ ion, X$^2\Sigma_g^+$, while the $o$ state is the first member
of the series converging on the first ionic excited state,
A$^2\Pi_u$. In the case of the $b^{\prime}$ and $b$ valence states,
the extended widths of the corresponding PECs in
Fig.~\ref{fig:N2potentials} are due to the aforementioned
configurational mixing. In addition, there are significant
electrostatic interactions within the manifolds of a given symmetry,
Rydberg-valence for $^1\Sigma_u^+$, and Rydberg-valence and
Rydberg-Rydberg for $^1\Pi_u$, since the MO configurations of all of
the isosymmetric states differ in exactly two of the occupied electron
orbitals \citep{lefebvre-brion_field2004}. The PECs in Fig. 1 are
shown in the diabatic (crossing) representation.

Most of the rovibrational levels of the singlet excited states are
predissociated, i.e., the molecule is initially bound following
photoabsorption, but then dissociates on timescales of a nanosecond or
less due to direct or indirect coupling to a dissociative continuum.
For the $^1\Pi_u$ states considered here, spin-orbit coupling to the
strongly-coupled and -predissociated $^3\Pi_u$ manifold (red PECs in
Fig.~\ref{fig:N2potentials}), with ultimate dissociation via the $C^{\prime}$ state,
provides the predissociation mechanism \citep{Lewis05b,lewis_etal2008b},
with a minor contribution from a crossing by the $2\,{^3\Sigma_u^+}$
state (green PEC in Fig.~\ref{fig:N2potentials}) at higher energies.  For the
$^1\Sigma_u^+$ states, two mechanisms are important \citep{heays2011thesis}:
first, a similar spin-orbit coupling to the $^3\Pi_u$ manifold, solely
responsible for predissociation in the absence of rotation, and
second, rotational coupling between the $^1\Sigma_u^+$ and $^1\Pi_u$
manifolds, followed by the $^1\Pi_u$ predissociation described above.
For the wavelengths considered here, as implied by Fig.~\ref{fig:N2potentials}, these
mechanisms result in primarily N$(^4S)+$N$(^2D)$ dissociation
products, i.e., one of the nitrogen atoms is formed in an excited
electronic state which decays on a timescale of 17~h into the ground
state N$(^4S)$. This is consistent with the observations of
\citet{Walter93} who failed to detect direct N$(^4S)+$N$(^4S)$
dissociation products.

\chg{The line-by-line models previously used to compute the N$_2$
  photodissociation rate require knowledge of the wavelengths,
  oscillator strengths, lifetimes, and predissociation probabilities
  of (transitions to) all rovibrational levels associated with the
  coupled excited singlet states.}
For the case of the
isoelectronic molecule CO, molecular models have been built previously
by specifying the term values, rotational and vibrational constants,
oscillator strengths, Einstein $A$ coefficients, and predissociation
probabilities for each excited electronic state
\citep[e.g.,][]{vanDishoeck88CO, Viala88, Lee96, Visser09}.  These
have allowed the rotationally-resolved absorption spectra of CO and
its isotopologues to be constructed using simple scaling relations.
Such models must be validated by a large quantity of laboratory data
and have been shown to be incorrect when strong interactions occur
between electronic states and their differing energetics.  For the
case of N$_2$, it is known that there are many wide-scale
perturbations, together with rapid dependences of oscillator strengths
and predissociation linewidths on rotational quantum number $J$, and
strong, irregular isotopic effects. It is impossible to fully
reproduce these effects using only a few spectroscopic constants.


The best way to simulate the N$_2$ spectrum, and the method employed
here, is, at each energy, to solve the full radial diabatic
coupled-channel Schr{\" o}dinger equation (CSE) for the coupled
electronic states described above, including all electrostatic,
spin-orbit, and rotational couplings, using the quantum-mechanical
methods of \citet{vanDishoeck84nonad}.  This is a {\em physically-based}
technique, with great predictive powers which enables confidence in the
computed spectrum in regions lacking experimental confirmation, \chg{even where perturbations are present}.
Furthermore, computations of isotopic spectra require only the change
of a single parameter, i.e., the reduced molecular mass, in the
molecular model: the results can be guaranteed since the underlying
physics is the same for all isotopologues. The same cannot be said for
the \chg{line-by-line} models such as those employed for CO, which would also
benefit from a CSE approach.  The detailed CSE model for N$_2$
employed here has been described in \citet{heays2011thesis},\footnote{Available on-line at \texttt{http://hdl.handle.net/1885/7360}} incorporates
earlier models of the $^1\Pi_u$ \citep{Lewis05b,
  haverd_etal2005} and $^3\Pi_u$ states \citep{lewis_etal2008b} and
has been tested extensively against laboratory data, including
high-resolution spectra obtained at the SOLEIL synchrotron facility
\citep{heays2011thesis,Heays11}.
\chg{A complete discussion of the CSE model
  and a full listing of computed spectroscopic data is deferred to
Heays et al.\ (in prep.). }

For a given
rotational-branch transition, combining the excited-state
coupled-channel wavefuction with the $X$-state radial wavefunction and
appropriate diabatic allowed transition-moment components yields the
corresponding \chg{(continuous with wavelength)} photoabsorption cross section, with the
computed linewidths providing the required predissociation lifetime
information.  Total cross sections for a given temperature, assuming
local thermodynamic equilibrium, are formed by summing the individual
branch cross sections, weighted by appropriate Boltzmann and H{\"
  o}nl-London factors, \chg{and including rotational levels with $J$ as high as 50}.

CSE photoabsorption cross sections, $\sigma_\mathrm{abs}$, are
computed here over the wavelength range 912--1000~\AA\ with a step size of
0.0001~\AA,\ and for temperatures of 10, 50, 100, 500, and 1000~K.  The
Doppler broadening of the spectral lines is taken into account by
convolution with a Gaussian profile having a thermal line width.

A 10\% uncertainty is estimated for the total magnitude of the
photoabsorption cross section and principally arises from the absolute
uncertainty of the calibrating laboratory
spectra \citep{haverd_etal2005,heays2011thesis}.  The laboratory
measurements in question were recorded at 300\,K or below, so the
uncertainty may be somewhat larger for calculations employing an
extrapolation to 1000\,K.  Additionally, 3\% of the 1000\,K ground
state population will be in the first vibrational level, leading to a
slight redistribution of the absorption cross section into hot bands.
This is considered in the model calculations.

Photodissociation cross sections, $\sigma_\mathrm{pd}$=$\eta\times
\sigma_\mathrm{abs}$, are obtained from the photoabsorption cross
sections by comparing the predissociation and radiative lifetimes for
each rovibrational level. The predissociation efficiency $\eta$ is then given
by $\eta=1-\tau_{\rm tot}/\tau_{\rm rad}$, where $\tau_{\rm tot}$ is
the inverse of the sum of the radiative and predissociation rates. For
almost all transitions, $\eta\simeq1$: significant corrections for
partial dissociation are needed only for the $b-X(1,0)$ and
$c^{\prime}-X(0,0)$ bands near 986 and 959\ \AA, respectively
\citep{Lewis05a,Liu08,sprengers_etal2004b,wu2011}.  For example, the
top panel of Fig.~\ref{fig:Partition-function} illustrates the
CSE-computed branching ratio between spontaneous emission back to the
ground state and dissociation as a function of rotational level for
$c^{\prime}-X(0,0)$.  Such calculations were performed for all bands
appearing between 955 and 991~\AA.  The difference between calculated
absorption and dissociation cross sections for the very-strongly
absorbing $c^{\prime}-X(0,0)$ band is demonstrated in
Fig.~\ref{fig:cross_section}, revealing a significant alteration of
the band profile once the dissociation efficiency is considered.

The bottom panel of Fig.~\ref{fig:Partition-function} shows the
thermal population for various $J$ levels of the ground vibrational
state, assuming several temperatures. By comparing this with the top panel
of Fig.~\ref{fig:Partition-function} it can be seen that the
dissociation fraction for this band will depend significantly on the
temperature.

\placefigpartitionfunction
\placefigcrossections

\subsection{Photodissociation rates}

The photodissociation rate, $k_\mathrm{pd}$, of N$_2$ exposed to UV radiation
can be calculated according to
\begin{equation}
 k_{\rm pd} = \int \sigma_{\rm pd} (\lambda) I(\lambda)d \lambda\ \ {\rm s^{-1}},  \label{eq:kpd}
\end{equation}
where the
photodissociation cross section, $\sigma_\mathrm{pd}$, is in units of
cm$^2$ and $I$ is the mean intensity of the radiation in photons
cm$^{-2}$\,s$^{-1}$\,\AA$^{-1}$\ as a function of wavelength,
$\lambda$, in units of \AA. The unattenuated interstellar radiation
field according to \citet{Draine78} is used in most of the following
calculations and is given by
\begin{equation}
 I(\lambda) = 3.2028 \times 10^{15} \lambda^{-3} - 5.1542 \times 10^{18} \lambda^{-4} + 2.0546 \times 10^{21} \lambda^{-5}\label{eq:draine field}.
\end{equation}

Inside a cloud, self-shielding, shielding by H, H$_2$, CO and other
molecules, and continuum shielding by dust all reduce the
photodissociation rate below its unattenuated value $k^0$. The
shielding function is defined to be
\begin{equation}
\Theta = k / k^0 \label{eq:shielding function}
\end{equation}
and can be split into a self-shielding,
\begin{equation}
\Theta_{SS} =  \frac{\int I(\lambda) \exp\left[-N({\rm N}_2) \sigma_{\rm abs}(\lambda)\right]\sigma_{\rm pd}(\lambda)\,d\lambda}{\int I(\lambda)\,\sigma_{\rm pd}(\lambda)\,d\lambda}, \label{eq:self shielding function}
\end{equation}
and a mutual-shielding part,
\begin{equation}
\Theta_{MS} =  \frac{\int I(\lambda)\exp\left[-N(X) \sigma_{\rm X}(\lambda)\right]\sigma_{\rm pd}(\lambda)\,d\lambda}{\int I(\lambda)\,\sigma_{\rm pd}(\lambda)\,d\lambda}  \label{eq:mutual shielding function}
\end{equation}
Here, $X$=H, H$_2$ or CO and $N$ is
the column density of the various species. A dust extinction term,
$\exp(-\gamma A_{\rm V})$, can be written in place of the exponential term
in Eq.~(6) where $A_{\rm V}$ is the optical depth in magnitudes and $\gamma$
depends on the assumed properties of the dust.  This is further
discussed in \S~\ref{sec:shielding}.  In all cases, the integrals
above are computed between 912 and 1000\,\AA.


\section{Results} \label{sec:results}

\subsection{Unattenuated interstellar rate} \label{sec:UIrate}

\placefigoverview

Figure~\ref{fig:N2overview} shows model spectra of N$_2$ and
${\rm{}H}_2+{\rm{}H}$ absorption for excitation temperatures of 50 and
1000~K.  At 50\,K the N$_2$ spectrum is made up of prominent
well-separated bands. These represent excitation to a range of
vibrational levels attributable to the five accessible electronic
states.  In contrast, the spectrum simulating a temperature of 1000\,K
includes the excitation of many more rotational levels and has few
sizable windows between bands.

Unshielded photodissociation rates of N$_2$ immersed in a
\citet{Draine78} field were calculated from the model
photodissociation cross section using Eqs.~(\ref{eq:kpd}) and
(\ref{eq:draine field}), and assuming a range of excitation
temperatures. These are plotted in Fig.~\ref{fig:rates} and listed in
Table~\ref{tab:BB-kpd0}.  The rate at 50~K is
$1.65\times{}10^{-10}$\,s$^{ -1}$, where the uncertainty of 10\% only
reflects the uncertainty in the cross sections, not the radiation
field (see below). This new value is 28\% lower than the value of
$2.30\times{}10^{-10}$\,s$^{-1}$ recommended by
\citet{vanDishoeck88}. The latter estimate was based on the best
available N$_2$ spectroscopy at the time, and has an
order-of-magnitude uncertainty. For comparison, the unshielded
photodissociation rate of N$_2$ at low $T$ is around 35\% smaller than
that of CO computed by \citet{Visser09}.

\placefigrates

Table ~\ref{tab:contribution} summarizes the contributions of
individual bands to the total unattenuated dissociation rate. It is
seen that the main contributions arise from bands 12, 21, 22, 23 and
24. Hence, the key wavelength ranges responsible for the
photodissociation of N$_2$ are around 940\,\AA\ and between 957--980\,\AA.

The calculated unattenuated rate of N$_2$ increases with increasing
temperature so that the value at 1000~K, $1.86\times
10^{-10}$\,s$^{-1}$, is 15\% higher than for 10~K.  This is largely
due to a variable but overall increase with rotational quantum number
$J$ of the photodissociation branching ratios of the $c'(v=0)$ and
$b(v=1)$ states.  This can be seen in
Fig.~\ref{fig:Partition-function} for the $c'(v'=0)$ state, where at
10~K all of the excited population is in levels with $J=0-3$.  These
levels have a low predissociation probability and hardly contribute
to the photodissociation rate.  At higher temperatures, the excited
population shifts to higher $J$, and at 1000~K the distribution
maximum occurs around $J=15-20$ for which the branching ratio to
dissociation is much higher.

\placetabBB

The rate obviously depends on the choice of radiation field. For
the alternative formulations of \citet{Habing68},
\citet{Gondhalekar80} and \citet{Mathis83}, the unattenuated rates are
$1.45$, $1.34$ and $1.51\times 10^{-10}$ s$^{-1}$ at 50~K,
respectively.  Additionally, Table~\ref{tab:BB-kpd0} considers the unattenuated
rates of N$_2$ assuming different blackbody radiation fields.  In
these calculations, the intensities have been normalized such that the
integrated values from 912--2050\,\AA\ are the same as those of the
\citet{Draine78} field.  The adopted dilution factors are $1.9\times
10^{-9}$, $3.4\times 10^{-12}$, $1.2\times 10^{-13}$, $1.6\times
10^{-14}$, and $1.6\times 10^{-16}$ for blackbody temperatures of
4000, 6000, 8000, 10\,000 and 20\,000\,K, respectively.  The value of
the unattenuated rate of N$_2$ at $4000$\,K (cool star) is $6$ orders
of magnitude smaller than that at $20\,000$\,K (hot star), and
increases steeply with stellar effective temperature. The
photodissociation rate of N$_2$ at $20\,000$\,K is comparable to that
in the Draine interstellar field, $1.65\times 10^{-10}$\,s$^{-1}$.
The calculated photodissociation rates for temperatures of 4000 and
$10\,000$ K are close to those recommended by \citet{vanDishoeck06}.

\placetabcontributions

\subsection{Self-shielding}
\placefigselfshielding

Although self-shielding is generally less important than mutual
shielding for the case of N$_2$ (see \S 3.3), it is potentially
important in protoplanetary disks and has been proposed to be
responsible for the enrichment of $^{15}$N in bulk chondrites and
terrestrial planets \citep{Lyons09,Lyons2010}. In this work, we compute the
self-shielding functions of N$_2$ at excitation temperatures of 10,
100 and 1000~K using the model absorption spectrum. Since this
spectrum is constructed using thermal line widths and no turbulent
broadening, it provides the maximum amount of shielding. For
reference, the thermal widths of N$_2$ at 10, 100 and 1000 K
correspond to full widths at half maximum of 0.1, 0.3 and 1.0 km s$^{-1}$.

As can be seen in Fig.~\ref{fig:self-shielding}, the
photodissociation of N$_2$ is free of self-shielding up to a column
density of around $10^{12}$\,cm$^{-2}$, but is fully shielded by
$10^{18}$ cm$^{-2}$. For intermediate N$_2$ column densities the
self-shielding function increases with increasing excitation
temperature. There are two reasons for this \citep{Visser09}.  First,
the optical depth of each line increases linearly with the thermal population
of its corresponding lower-state rotational level, but the
self-shielding increases nonlinearly according to Eq.~(\ref{eq:self
  shielding function}).  Then, because the ground state population at
higher temperatures is distributed over more levels (see
Fig.~\ref{fig:Partition-function}) there is an overall decrease in the
effectiveness of self-shielding.  The second effect arises from the
individual line profiles, which are constructed to have thermal
broadening.  Those lines appearing in the 1000\,K spectrum are then 10
times broader than those at 10\,K, leading to decreased peak optical
depth at the line center and less effective self-shielding over the
whole line profile.

\subsection{Shielding by H$_2$, H and CO} \label{sec:shielding}

\placefigCOspectra

The wavelength range over which N$_2$ can be photodissociated is
exactly the same range over which H$_2$, H and CO absorb strongly.
The amount by which N$_2$ is shielded depends on the column densities
of each of these species and is characterized by the shielding
function of Eq.~(\ref{eq:mutual shielding function}).

Figure \ref{fig:N2overview} overlays absorption spectra for N$_2$ and
H+H$_2$ combined.  Two forms of the latter are included: \chg{a
  representative example spectrum deduced from observations of
  H$_2$($J$) and H column densities of the well-studied and
  commonly-referenced diffuse cloud toward $\zeta$ Oph; and a
  simulated spectrum using column densities of $2.1\times 10^{20}$ and
  $2.6\times 10^{20}\,{\rm cm}^{-2}$ for H$_2$ and H, respectively,
  and assuming purely thermal excitation of H$_2$.}  The H$_2$
molecular data adopted for the synthetic spectra are those of
\citet{Abgrall93a,Abgrall93b} and were obtained from the Meudon PDR
code website \citep{LePetit06}.  The assumed column densities were
taken to be half those of the observed $\zeta$ Oph cloud, as is
appropriate for radiation penetrating to its center, and an excitation
temperature of 50\,K was used for the H$_2$+H and N$_2$ thermal
models.  The principal difference between observed and thermal H$_2$+H
spectra is the appearance of additional lines in the observed spectrum
from non-thermally populated higher-$J$ levels.  Thermal excitation
H$_2$ spectra are used throughout the following mutual-shielding
calculations, and do not include extrathermal excitations such as UV
pumping.  This negligence leads to a slight (approximately 3\%)
underestimate of shielding for the case of the $\zeta$ Oph cloud.  A
magnified version of the spectra in Fig.~\ref{fig:N2overview} is
included in the online appendix, and it is apparent that the ranges
containing significant N$_2$ absorption and minimal shielding by H and
H$_2$ are 919.8--920.2, 921.2--921.6, 922.6--923.1, 925.8--926.1,
935.1-935.4, 939.9--940.3, 942.3--942.8, 958.1--958.9, 959.0--959.1,
960.1--960.8 and 978.8--979.5\,\AA.

The calculated N$_2$ photodissociation rate at the centre of the
$\zeta$ Oph cloud is $6.96\times 10^{-11}$\,s$^{-1}$, corresponding to
58\% shielding by H$_2$+H.  Table~\ref{tab:contribution} summarizes
the contributions to the photodissociation rate of individual bands at
the edge of the cloud (unshielded) and at its center.  The pattern
of increasing and decreasing significance of individual N$_2$ bands
under the influence of shielding is easily matched to the occurrence
of overlapping features in Fig.~\ref{fig:N2overview}(a).  The heavy
shielding of bands 22 and 23 has a particularly large effect on the
total photodissociation rate, the relative importance of the lightly
shielded band 14 increases significantly in the center, and the
$957-980$\,\AA\ wavelength range remains particularly important for
photodissociation throughout the cloud.

A similar investigation was performed considering the shielding of
N$_2$ by CO.  Simulated absorption spectra for both molecules are
shown in Fig.~\ref{fig:N2COspectra}, where the CO spectrum was
generated by the photoabsorption model of \citet{Visser09} assuming a
column density of $10^{15}$\,cm$^{-2}$, close to half of the observed
$\zeta$ Oph value.  Both spectra exhibit a complex pattern of bands so
that overlaps are infrequent and do not occur at all in the most
important photodissociation range, 957--980\,\AA{}.  In this range CO
hardly affects N$_2$ and, in general, shielding by H$_2$ and H is
sufficiently dominant that the additional influence of CO can be
neglected.

\placetabshielding
\placefigshielding

Two-dimensional shielding functions for a range of H$_2$ and H column
densities have been calculated.  These are tabulated in
Table ~\ref{tab:shielding-logN(H)=14,20,22} and shown graphically in
Fig.~\ref{fig:shielding-diff-logN(H)}.  For these calculations an
excitation temperature of 50\,K was assumed for both N$_2$ and H$_2$,
and $b({\rm H}_2)$ (the nonthermal broadening) was set to
3\,km\,s$^{-1}$.  Obviously, the shielding function decreases with
increasing $N$(H$_2$) and $N$(H), but H$_2$ plays the more important
role (as is the case for the shielding of CO; \citealt{Visser09}).
Specifically, N$_2$ is close to fully shielded ($\Theta <2\%$) when
$N({\rm H}_2)=10^{22}\,{\rm cm}^{-2}$, and totally shielded by
$10^{23}$\,cm$^{-2}$.  Electronic tables of the calculated shielding
functions can be obtained from {\tt
  www.strw.leidenuniv.nl/$\sim$ewine/photo}.

\placefigshieldingtemperatures

Since N$_2$ does not possess a permanent dipole moment, radiative
decay from excited rotational levels of its electronic-vibrational
ground state is slow.  Then, the excitation temperature of N$_2$ is
likely to be higher than that of CO and other molecules, and closer to
the kinetic temperature.  The effect of temperature on shielding by
H$_2$+H was investigated and is illustrated in
Fig.~\ref{fig:shieldingtemperatures}.  The same excitation
temperatures are adopted for H$_2$ and N$_2$ because both are
zero-dipole-moment molecules.  The calculated shielding functions are
somewhat erratic, and even show a peculiar non-monotonic temperature
dependence at low H$_2$ column density. This arises from the small
degree of overlap occurring between atomic H lines and N$_2$ bands.  The
distribution of N$_2$ lines over additional rotational transitions at
higher temperatures leads to the variability of
Fig.~\ref{fig:shieldingtemperatures} and illustrates the need for
high-resolution reference spectra in these kinds of applications.
For significant H$_2$ column densities, and in contrast with N$_2$
self-shielding, the amount of shielding increases with increasing
temperature.  This results from a H$_2$ population that is spread over
more rotational levels at higher temperatures, leading to an
absorption spectrum featuring more lines available to shield
N$_2$. This is clearly evident when comparing the various curves in
Fig.~\ref{fig:N2overview}.

\placetabcodust

Table~\ref{tab:shielding-co-N2-dust} compares the H+H$_2$ shielding of
N$_2$ with the CO shielding calculations of \citet{Visser09}.  The two
molecules follow a similar pattern, within 50\%, up to $N({\rm
  H}_2)=10^{22}$\,cm$^{-2}$.  This difference becomes more significant
when $N({\rm H}_2)=10^{23}$\,cm$^{-2}$, but photodissociation has long
ceased to be important as an N$_2$ destruction mechanism by then.

\subsection{Shielding by dust}

Dust grains compete with molecules in the cloud by also absorbing UV
photons. For the 912--1000 \AA\ wavelength range, the attenuation by
dust is largely independent of wavelength and can be taken into
account by an additional shielding term $\exp(-\gamma A_{\rm V})$
\citep{vanDishoeck06}. For the wavelength range appropriate for N$_2$,
a value of $\gamma=3.53$ is \chg{found, using the method of
  \citet{Roberge81} and standard diffuse-cloud grain properties of
  \citet{Roberge91}}. For larger dust grains of a few $\mu$m in size,
such as is appropriate for protoplanetary disks, $\gamma \approx 0.6$
\citep{vanDishoeck06}. The visual extinction $A_{\rm V}$ is computed
from the total hydrogen column $N_{\mathrm{H}}$=$N$(H) + 2$N$(H$_2$)
through the relation $A_{\mathrm{V}}=N_{\mathrm{H}}/1.6\times
10^{21}$, based on \citet{Savage77}.

\chg{For diffuse clouds with total visual extinctions around 1 mag,
  radiation from the other side of the cloud may result in a shallower
  depth dependence than given by the above single exponential form. In
  such cases, a bi-exponential form $\exp\left({-\alpha A_{\rm V} + \beta
    A_{\rm V}^2}\right)$ may be more appropriate
  \citep{vanDishoeck84,vanDishoeck88}. For $A_{V}^{tot}$=1 mag,
  $\alpha=7.25$ and $\beta=6.92$ are found.}

The shielding of N$_2$ by dust under various conditions is listed in
Table~\ref{tab:shielding-co-N2-dust}. This shows that shielding by
normal interstellar dust is larger than that by H$_2$ and H for any
$A_{\rm V}$ and implies that the `smoke screen' by dust also plays a
significant role in diffuse and translucent clouds and
photon-dominated regions. However, in protoplanetary disks where the
larger dust particles absorb and scatter less efficiently, the effects
of H$_2$ and H shielding become comparable, or even dominant, at large
$A_{\rm V}$.


\section{Chemical models}

As an example of how to apply the new photodissociation rates, we ran
chemical models for a set of diffuse and translucent clouds, a
photon-dominated region (PDR), and a vertical cut through a
circumstellar disk. The models use the UMIST06 chemical network
\citep{Woodall07}, stripped down to species containing only H, He, C,
N and O. Species containing more than two C, N or O atoms are also
removed since they are not relevant for our purposes. Freeze-out and
thermal evaporation are added for all neutral species, but no
grain-surface reactions are included other than H$_2$ formation
according to \citet{Black87}. Self-shielding of CO is computed using
the shielding functions of \citet{Visser09}; for N$_2$, we use the
self-shielding functions calculated here at 50\,K. The elemental
abundances relative to H are $0.0975$ for He, $7.86\times10^{-5}$ for
C, $2.47\times10^{-5}$ for N and $1.80\times10^{-4}$ for O
\citep{Aikawa08}.  Enhanced formation of CH$^+$ (and thus also CO) at
low $A_{\rm V}$ is included following \citet{Visser09} by
supra-thermal chemistry, boosting the rate of ion-neutral reactions by
setting the Alfv{\'e}n speed to 3.3\,km\,s$^ {-1}$ for column
densities less than $4\times 10^{20}$ cm$^{-2}$.  Unless stated
otherwise, the model of impinging UV flux is the Draine field of
Eq.~(\ref{eq:draine field}) modified by a scaling factor, $\chi$. In
all cases, the abundances of N, N$_2$ and CO reach steady state after
$\sim$1 Myr, regardless of whether the gas starts in atomic or
molecular form.

\subsection{Diffuse and translucent clouds}


\placefigtrans
\placefigcoldentransgrid

A set of diffuse and translucent cloud models was run for central
densities $n_{\rm H}=n({\rm{}H})+2n({\rm{}H}_2)=100, 300$ and $10^3$
cm$^{-3}$, at a temperature of 30 K, and assuming scaling factors of
the UV flux of $\chi=0.5, 1$ and 5. Figure~\ref{fig:trans} shows the
abundances of N, N$_2$ and CO and the photodissociation rates of N$_2$
and CO as functions of depth into the cloud (measured in $A_{\rm V}$)
for the $n_{\mathrm{H}}=10^3$ and $\chi$=1 model. Both CO and N$_2$ are rapidly
photodissociated in the limit of low extinction and carbon and
nitrogen are primarily in atomic form.  Some CO is formed in a series
of (supra-thermal) ion-molecule reactions starting with C$^+$ at the
edge \citep{Visser09}. Since the ionization potential of atomic N lies
just above that of H, preventing the formation of N$^+$, N$_2$ can
only form through slower neutral-neutral reactions.
As a result, the abundance of N$_2$ is three orders of magnitude lower
than that of CO at the edge of the cloud.  The conversion from N to
N$_2$ occurs at an $A_{\rm V}$ of 1.5\,mag, at which point CO has
become the main form of carbon. The bottom panel of
Figure~\ref{fig:trans} illustrates that self-shielding and mutual
shielding by H and H$_2$ significantly reduce the
photodissociation rate relative to dust alone. The column densities
of N$_2$ and CO at $A_{\rm V}$=1.5\,mag are $1.5\times10^{15}$ and
$1.3\times10^{16}$ cm$^{-2}$. At high $A_{\rm V}$, atomic N is
maintained at an abundance of $3\times10^{-6}$ by the dissociative
recombination of N$_2$H$^+$, which in turn is formed from the reaction
between N$_2$ and cosmic-ray-produced H$_3^+$.

To investigate the role of turbulence or non-thermal motions on the
results, a model has been run in which the Doppler width of the N$_2$
lines in the self-shielding calculation was increased to
3\,km\,s$^{-1}$ rather than the thermal width at low temperatures. The
resulting N$_2$ abundance as a function of depth is nearly identical
to that presented in Fig.~\ref{fig:trans}.


Absorption bands of N$_2$ have possibly been detected in observations
of the diffuse cloud toward HD~124314 \citep{Knauth04}. \chg{The
  two relevant bands, indicated in Fig.~\ref{fig:N2overview},
  are particularly strongly absorbing, and are relatively unshielded by
  hydrogen.} The depth of the observed absorption indicates a total N$_2$
column density of $(4.6\pm\,0.8)\times 10^{13}$\,cm$^{-2}$ and the
stellar reddening of HD~124314 provides an estimate of the cloud's
extinction, $A_{\rm V}=1.5$\,mag.
Figure~\ref{fig:coldentransgrid} shows the cumulative N$_2$ column
density calculated for a range of radiation field intensities and
$n_\mathrm{H}$ densities as a function of $A_{\rm V}$.  For comparison
with the model, which considers only half of the cloud from edge to
center and is irradiated from one side only, the observed $A_{\rm V}$
and N$_2$ column density must be halved. \chg{These models use the
  single exponential dust continuum shielding function; if the
  bi-exponential formulation were used, the model N$_2$ column
  densities would be even lower for small $A_{\rm V}$.} The maximum
calculated column density occurs where the radiation field is weakest
($\chi=0.5$) and for the highest density
($n_\mathrm{H}=10^3\,\mathrm{cm}^{-3})$.  These are extreme physical
conditions for a cloud like HD 124314 and \chg{inconsistent with its
  relatively high H/H$_2$ column density ratio \citep{Andre03} and low
  CO column density \citep{Sheffer08}.} An independent conformation of
the N$_2$ detection is warranted.  Observed upper limits toward other
diffuse clouds with lower $A_{\rm V}$ are a few$\times
10^{12}$\,cm$^{-2}$ \citep{Lutz79}, which are consistent with the
current models for typical densities of a few hundred cm$^{-3}$ and
$\chi \geq$1.

\subsection{Photon-dominated region}
The PDR model is run assuming an $n_{\rm H}$ density of
$10^5$\,cm$^{-3}$, a temperature of 100 K, and a UV flux of
$\chi=10^3$. Figure~\ref{fig:pdr} shows the resulting abundances of N,
N$_2$, C, C$^+$ and CO and the relevant photodissociation rates as functions of $A_{\rm V}$.

\placefigpdr

The calculated abundances of both N$_2$ and CO at low $A_{\rm V}$ are
lower in the PDR model compared with the diffuse and translucent
clouds, because of the stronger UV field.  For all models, the
abundance of N$_2$ is several orders of magnitude lower than that of
CO.  Also, because of the increased radiation, the transition from N
to N$_2$ occurs deeper into the PDR: at an $A_{\rm V}$ of about 3
mag. The column densities of N$_2$ and CO at this point are
$6\times10^{15}$ and $3\times10^{16}$ cm$^{-2}$.  The minor wiggle in
the atomic N abundance profiles at $A_{\rm V}=2$--3 mag is due to the
abundance patterns of CH and OH. CH is the main destroyer of N at
$A_{\rm V}=2$ mag but its abundance drops going into the cloud because
its main precursor, C$^+$, disappears. The simultaneously increasing
extinction allows for an increase in the OH abundance, so that this
becomes the main destroyer of N for $A_{\rm V}>3$ mag. Interestingly,
the transition from N$\to$N$_2$ occurs at nearly the same depth into
the cloud than that of C$^+ \to$C$\to$CO.

\subsection{Circumstellar disk}

The fourth model simulates a vertical slice through a circumstellar
disk and its setup is identical to that of \citet{Visser09}. The slice
is located at a radius of 105 AU in the standard model of
\citet{Alessio99}, which supposes a disk of 0.07 $M_\odot$ and 400 AU
radius surrounding a T Tauri star of 0.5 $M_\odot$ and 2 $R_\odot$
radius. The surface of the slice, at a height of 120 AU, is
illuminated by the Draine (1978) field with $\chi=516$. We ran the model
assuming dust grain sizes of 0.1 and 1 $\mu$m. The results are plotted
in Fig.~\ref{fig:disk}.

\placefigdisk

Starting from the disk surface (high $z$) and moving inwards, the
abundance profiles of N, N$_2$ and CO show the same qualitative trends
as they do for the translucent cloud and PDR models. The main
difference arises from freeze-out of N$_2$ and CO for $z$ below
25\,AU, at which point the dust temperature drops below $\sim$20
K. The depletion of N$_2$ also drives down the abundance of N$_2$H$^+$
which, in turn, restricts the abundance of atomic N. In the models
presented here, nitrogen is fully converted into N$_2$ at heights
where no freeze-out occurs.

Increasing the grain size from 0.1 to 1 $\mu$m allows the UV field to
penetrate to heights of about 35 instead of 45 AU. The abundances of
N$_2$ and CO are then a factor of 10--100 lower in the intervening
zone, while that of atomic N is a factor of a few higher. Since atomic
N is a prerequisite for an active nitrogen chemistry leading to
species like HCN, this result illustrates that larger column densities
of nitrogen-containing molecules can be expected in disks with grain
growth.

The column densities integrated from surface to midplane are
$2\times10^{17}$ cm$^{-2}$ for N$_2$ and $1.2\times10^{18}$ cm$^{-2}$
for CO, regardless of the grain size, since the bulk of the N$_2$ and
CO are at high densities where UV photodissociation is negligible. The
total column of N is $5\times10^{16}$ cm$^{-2}$ for 0.1 $\mu$m grains
and $8\times10^{16}$ cm$^{-2}$ for 1 $\mu$m grains.

The models discussed above use a scaled \citet{Draine78} radiation
field. A much cooler radiation field was also considered by assuming a
4000\,K blackbody source scaled to the same flux between 912 and 2050\,\AA.  Then, both carbon and nitrogen are fully
molecular at the disk surface since photodissociation of both CO and
N$_2$ is negligible.  A smaller amount of atomic N is maintained by
chemical reactions for active nitrogen chemistry.
This result is qualitatively consistent with the
observation of \citet{Pascucci09} that the color of the radiation
field affects the nitrogen chemistry, although that study applied to
the inner rather than the outer disk.

\section{Conclusions} \label{sec:conclusions}

In this work, we compute accurate N$_2$ photodissociation rates in the
interstellar medium for the first time by employing new molecular data
on its electronic transitions.
The calculated N$_2$ photodissociation rate in an unattenuated
interstellar radiation field is $1.65\times 10^{-10}$\,s$^{-1}$
(50~K), $\sim$28\% lower than the previously recommended
value. This rate increases somewhat with temperature due to
$J-$dependent predissociation rates.

The simulated spectra reveal that the most important range for
photodissociation is 957--980 \AA, where H$_2$ and H absorption
significantly overlap with N$_2$ absorption. In contrast, CO only
weakly shields N$_2$. Self-shielding and mutual shielding functions
have been computed for a range of N$_2$, H$_2$ and H column densities.
For interstellar grains, shielding by dust is also effective.
In protoplanetary disks, where dust particles have grown to
$\mu$m size, the dust shielding becomes less than that of
H$_2$.

The new rates have been incorporated into models of diffuse and
translucent clouds, of a dense PDR, and a protoplanetary disk.  The
translucent cloud models show that the observed column of interstellar
N$_2$ in a translucent cloud with $A_{\rm V}=1.5$ mag can only be reproduced
if the density $n_\mathrm{H}$ is higher than $1000\,\mathrm{cm}^{-3}$
and the radiation field has an intensity of less than half of the
\citet{Draine78} field. For dense PDRs, the N$_2$ abundance only
becomes significant at extinctions of more than 3 mag into the cloud
but the transition of N $\to$ N$_2$ occurs at nearly the same depth as
that of $\mathrm{C}^+\to\mathrm{C}\to\mathrm{CO}$. Disk models show
that nitrogen is fully converted into N$_2$ at heights before
freeze-out occurs, irrespective of grain size. However, an active
nitrogen chemistry can take place in the upper layers of a disk where
not all nitrogen is locked up in N$_2$, except for very cool radiation
fields.  Altogether, data are now available to accurately model N$_2$
photodissociation in a wide variety of interstellar and circumstellar
media.

\begin{acknowledgements}
  Astrochemistry in Leiden is supported by the Netherlands Research
  School for Astronomy (NOVA), by a Spinoza grant and grant
  648.000.002 from the Netherlands Organisation for Scientific
  Research (NWO), and by the European Community's Seventh Framework
  Programme FP7/2007-2013 under grant agreements 291141 (CHEMPLAN) and
  238258 (LASSIE).  Calculations of the N$_2$ photodissociation cross
  sections were supported by the Australian Research Council Discovery
  Program, through Grant Nos. DP0558962 and DP0773050.

\end{acknowledgements}
\bibliographystyle{aa}





\Online

\begin{appendix}
\section{self-shielding functions}
In this section, we provide a table containing the self-shielding functions of N$_2$ at 10, 100 and 1000~K (data for Fig.~\ref{fig:self-shielding}).

\placetabselfshielding

\label{sec:appendix-tabself-shielding}
\end{appendix}

\begin{appendix}
\section{High resolution spectra}
\label{sec:appendix-spectra}

In this section, we provide the high resolution spectra of N$_2$ at 50~K (zoom in for Fig.~\ref{fig:N2overview}). The H$_2$ and H column densities are the same as those used in
Figure~\ref{fig:N2overview}.


\setcounter{figure}{0}
\begin{figure*}[htb]
\centering
\includegraphics[angle=0,width=0.9\hsize,height=5.4cm]{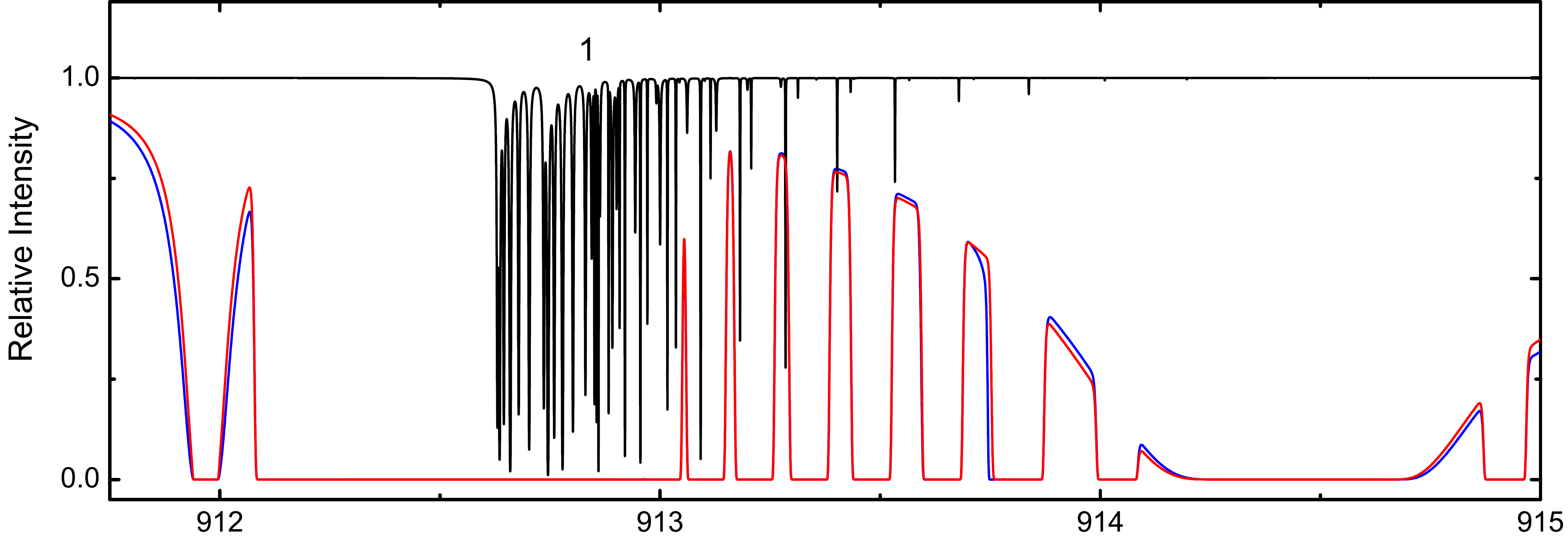}
\includegraphics[angle=0,width=0.9\hsize,height=5.4cm]{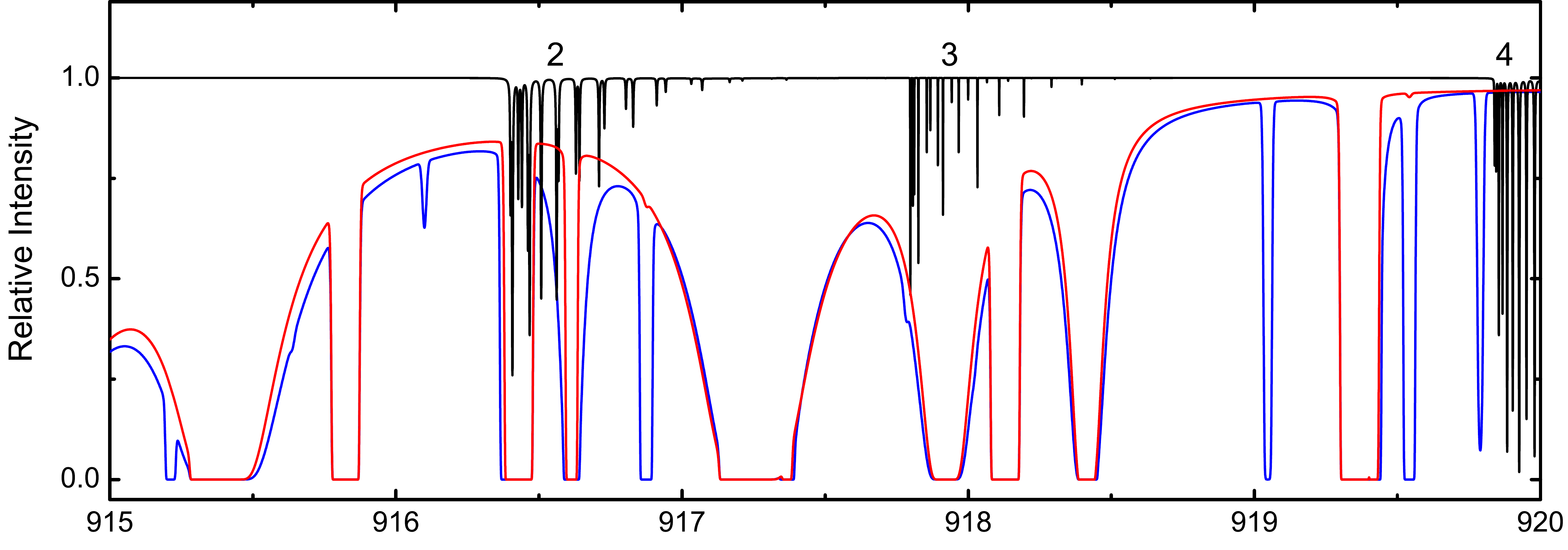}
\includegraphics[angle=0,width=0.9\hsize,height=5.4cm]{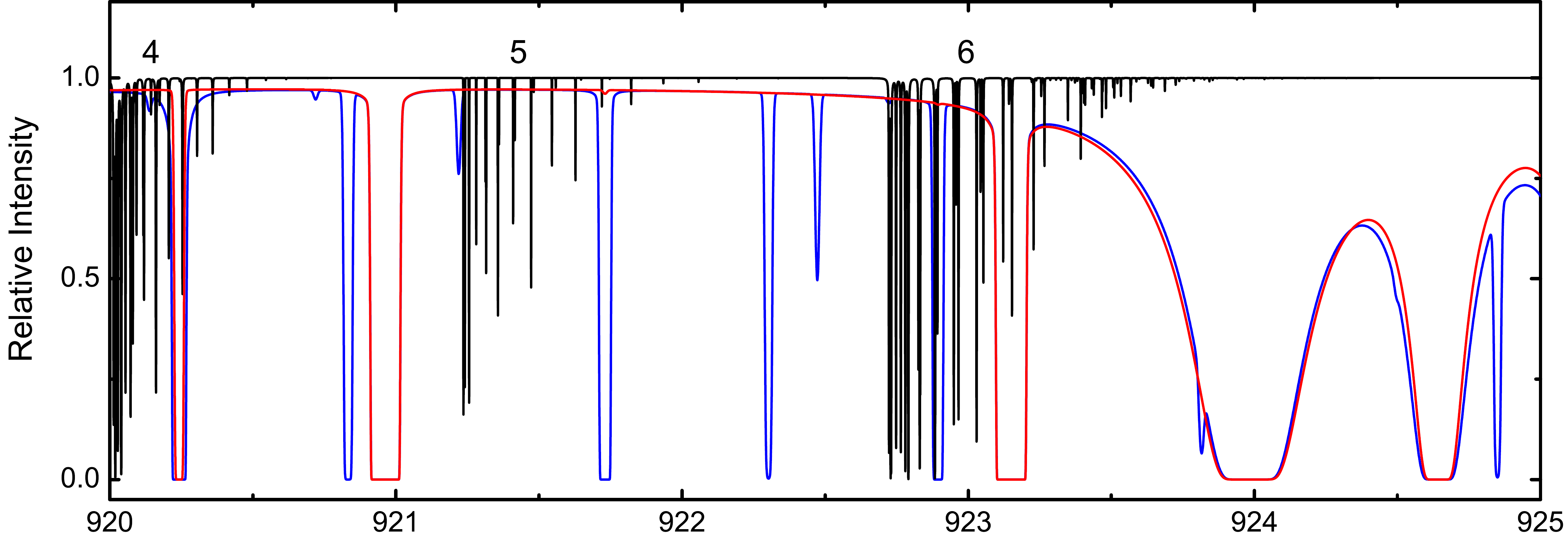}
\includegraphics[angle=0,width=0.9\hsize,height=5.4cm]{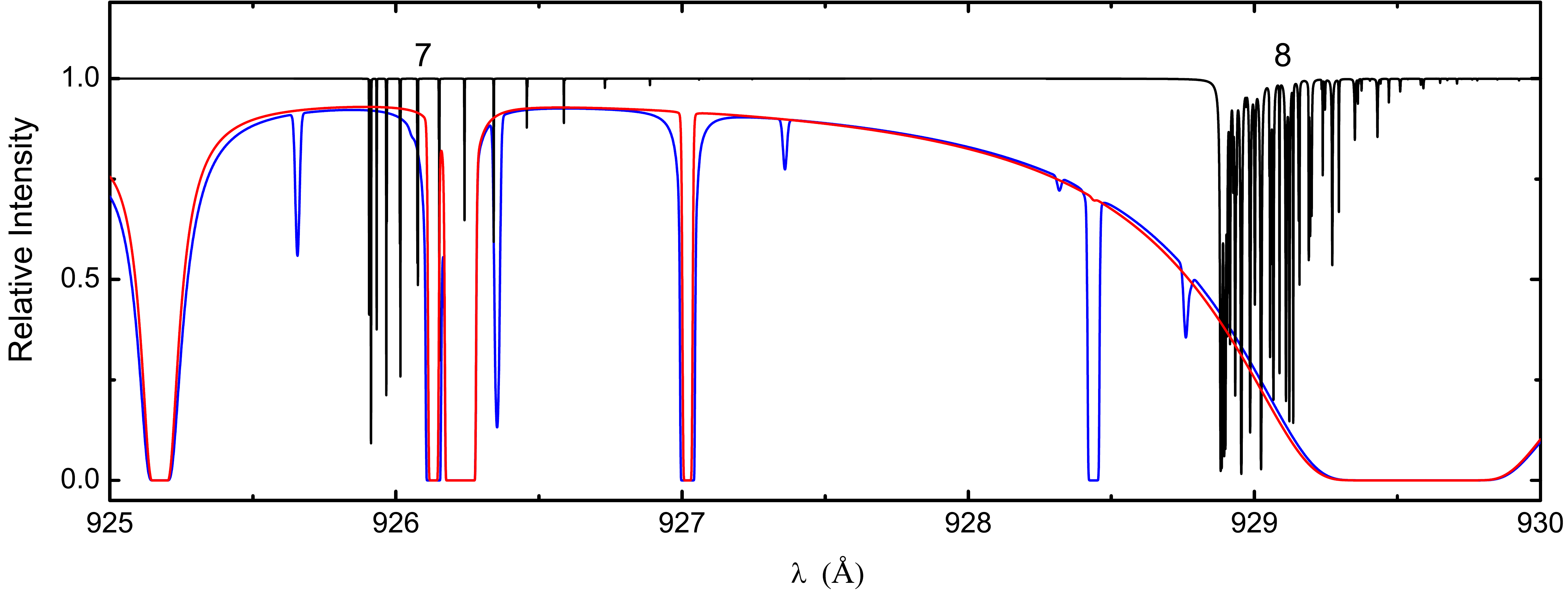}
\caption{Zoom-in of the high-resolution spectra of N$_2$ (black line) and H plus
  H$_2$ (red line) in the wavelength range of 911.75--930\,\AA{} for a
  thermal excitation temperature of 50~K.  The column density of N$_2$
  is $10^{15}$\,cm$^{-2}$ and values for H$_2$ and H are taken to be
  half of the observed column densities in the well-studied diffuse
  cloud toward $\zeta$ Oph, as is appropriate for the center of the
  cloud: $N$(H$_2$)=$2.1\times 10^{20}$ and
  $N$(H)=$2.6\times10^{20}$\,cm$^{-2}$. The model H$_2$ Doppler width
  is $3$\,km\,s$^{-1}$. Also shown is the ${\rm{}H}_2+{\rm{}H}$
  absorption spectrum (blue) towards $\zeta$ Oph using the observed
  column densities for individual $J$ levels, showing non-thermal
  excitation of H$_2$.}
\label{fig:N2detail1}
\end{figure*}

\setcounter{figure}{1}
\begin{figure*}[htb]
\centering
\includegraphics[angle=0,width=0.9\hsize,height=5.4cm]{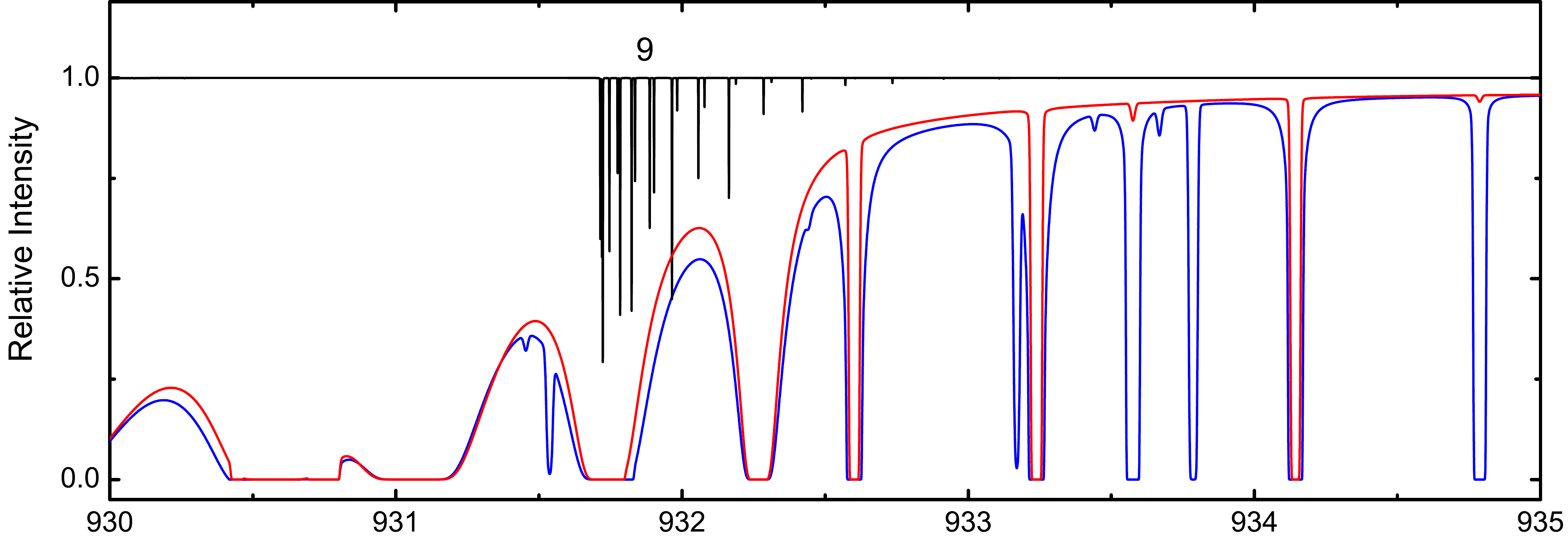}
\includegraphics[angle=0,width=0.9\hsize,height=5.4cm]{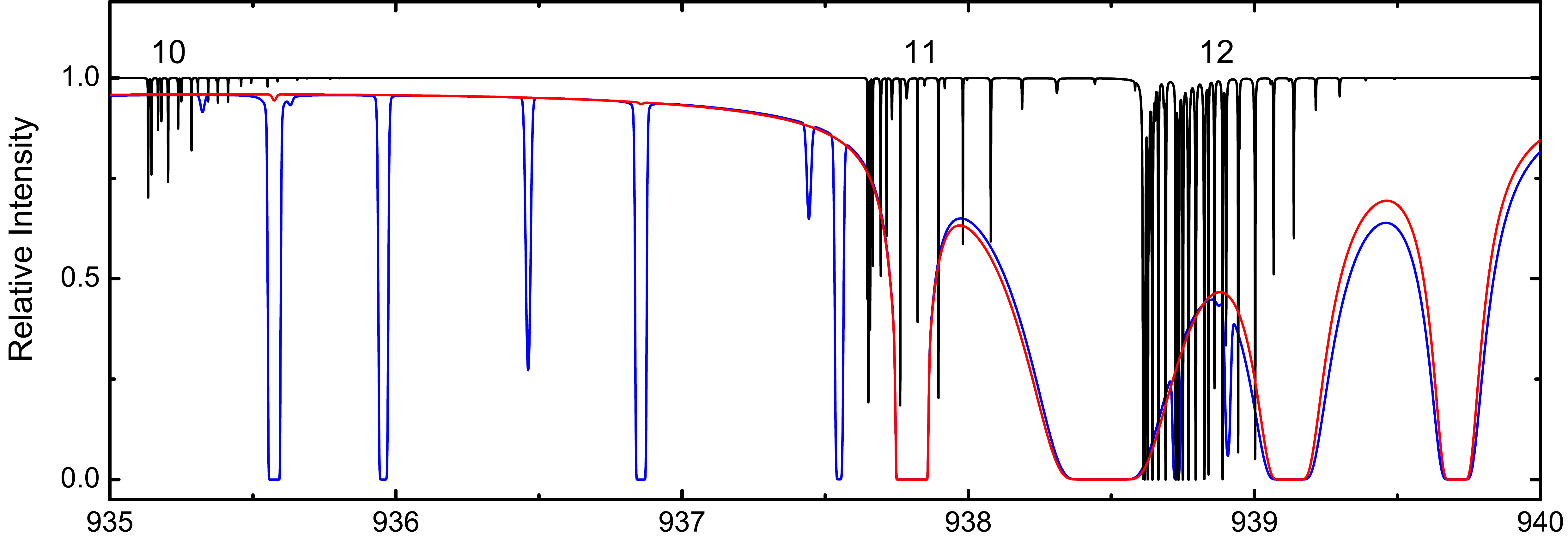}
\includegraphics[angle=0,width=0.9\hsize,height=5.4cm]{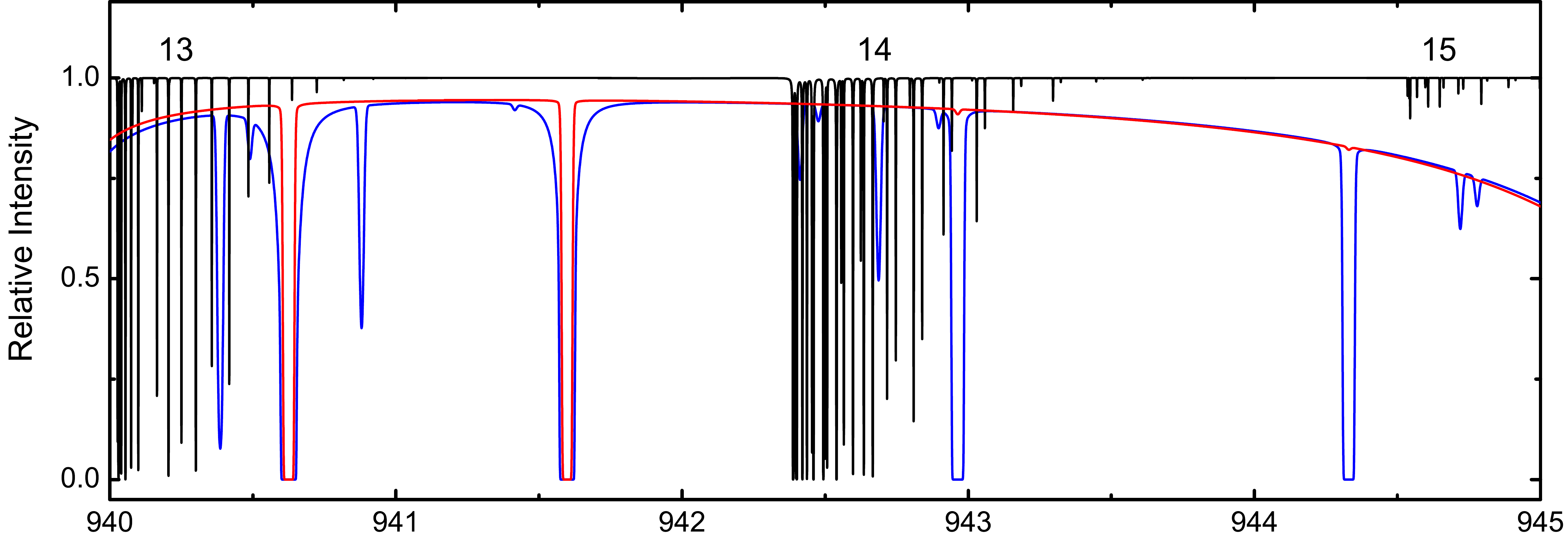}
\includegraphics[angle=0,width=0.9\hsize,height=5.4cm]{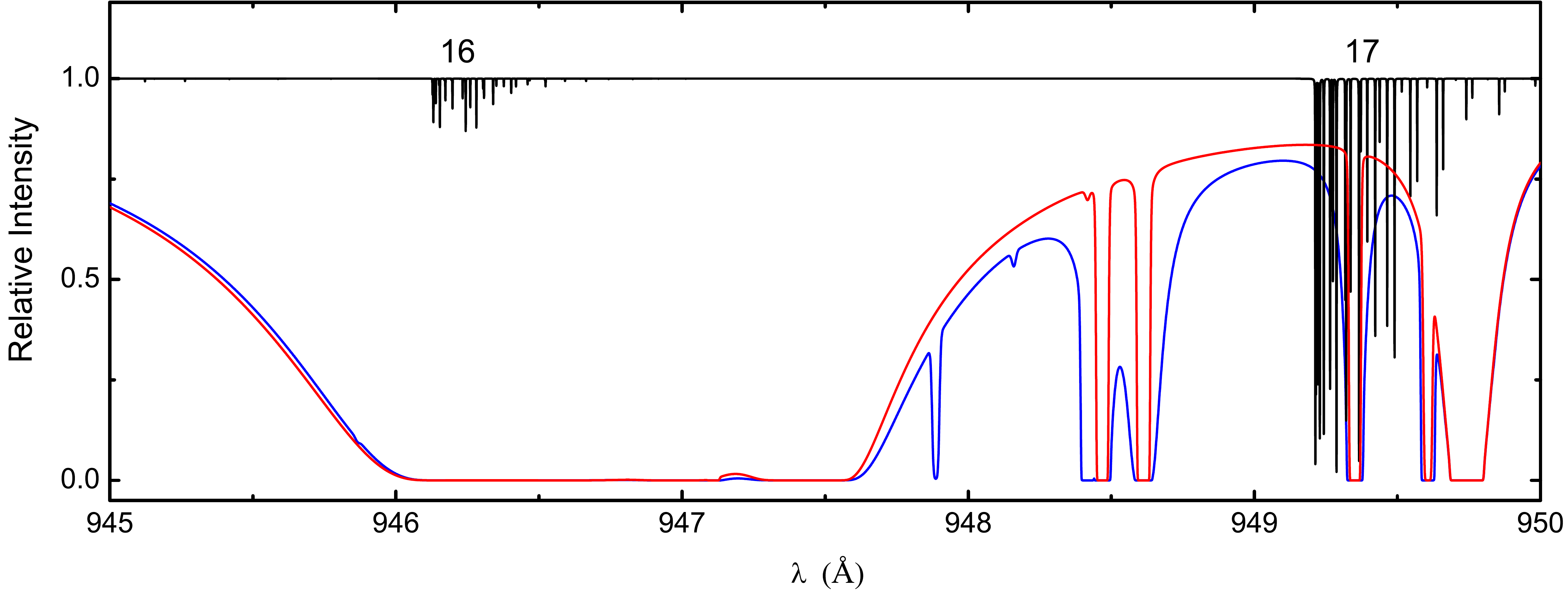}
\caption{As Fig.~\ref{fig:N2detail1}, but for 930--950 \AA.}
\label{fig:N2detail2}
\end{figure*}
\clearpage

\setcounter{figure}{2}
\begin{figure*}[htb]
\centering
\includegraphics[angle=0,width=0.9\hsize,height=5.4cm]{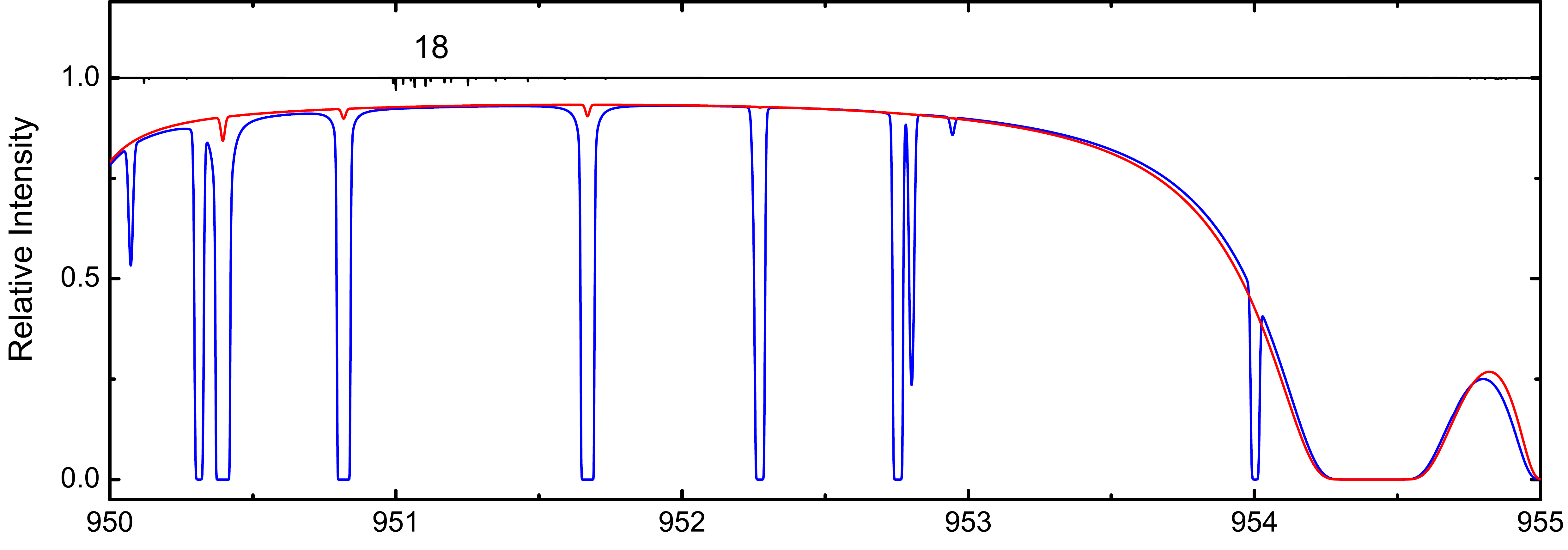}
\includegraphics[angle=0,width=0.9\hsize,height=5.4cm]{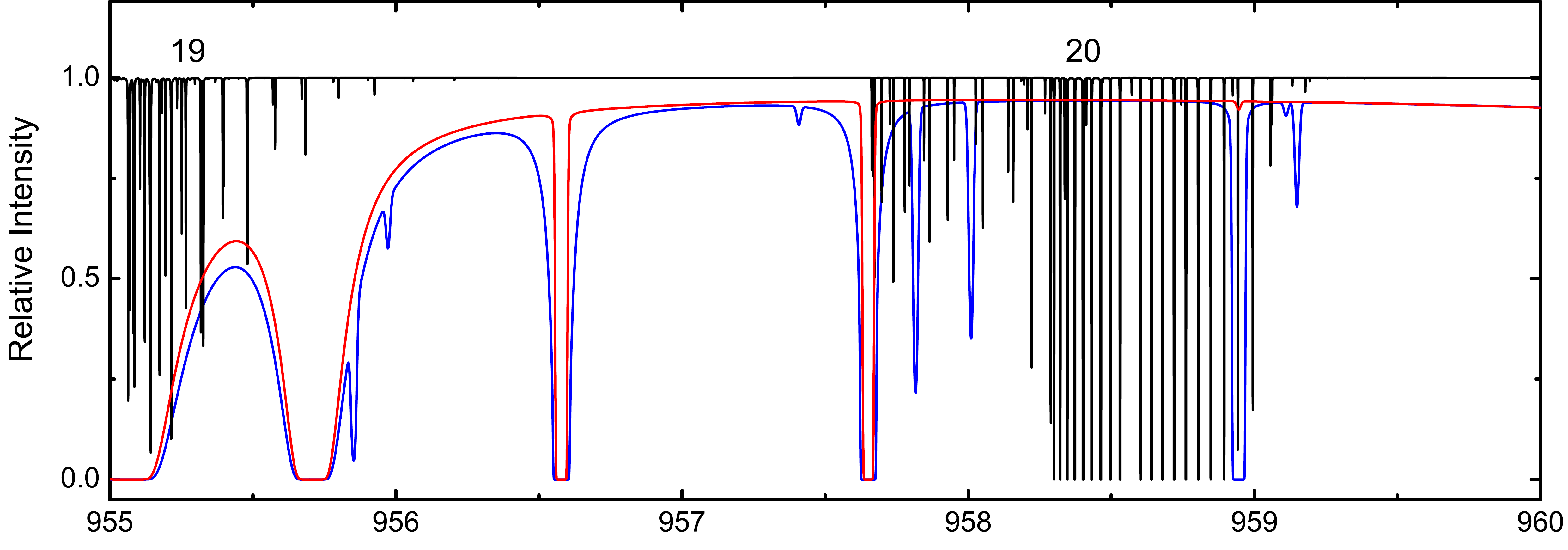}
\includegraphics[angle=0,width=0.9\hsize,height=5.4cm]{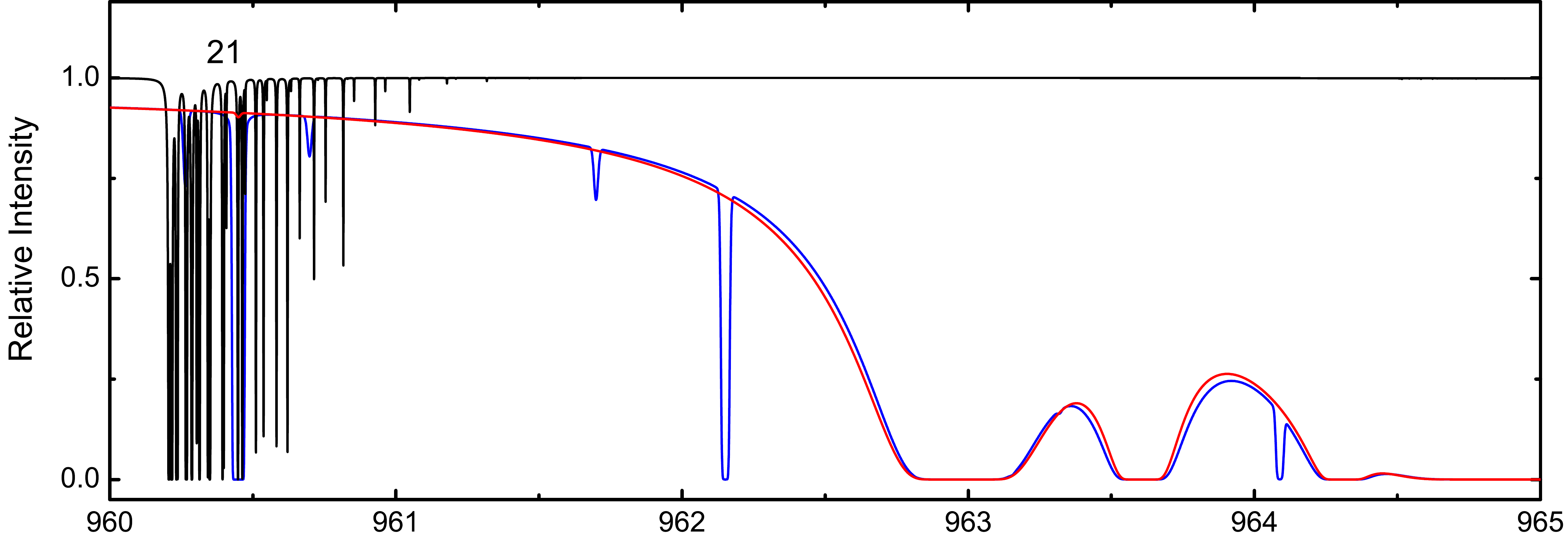}
\includegraphics[angle=0,width=0.9\hsize,height=5.4cm]{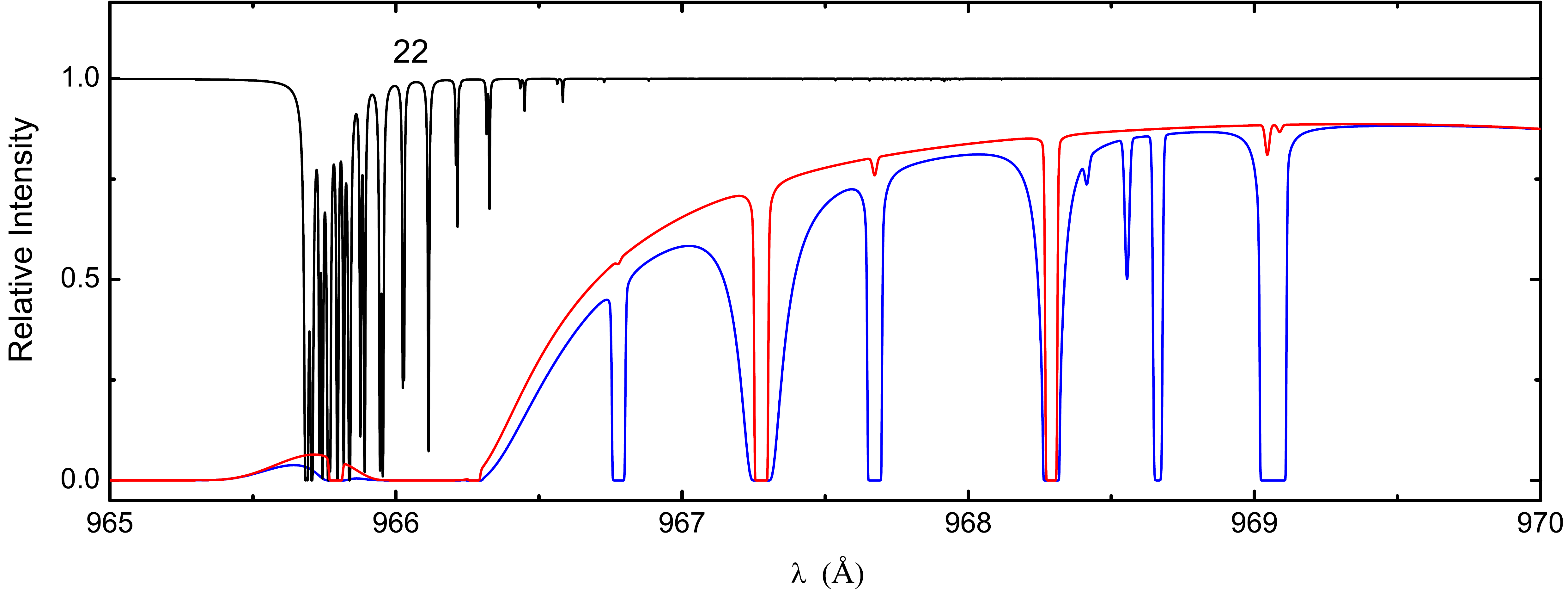}
\caption{As Fig.~\ref{fig:N2detail1}, but for 950--970 \AA.}
\label{fig:N2detail3}
\end{figure*}
\clearpage

\setcounter{figure}{3}
\begin{figure*}[htb]
\centering
\includegraphics[angle=0,width=0.9\hsize,height=5.4cm]{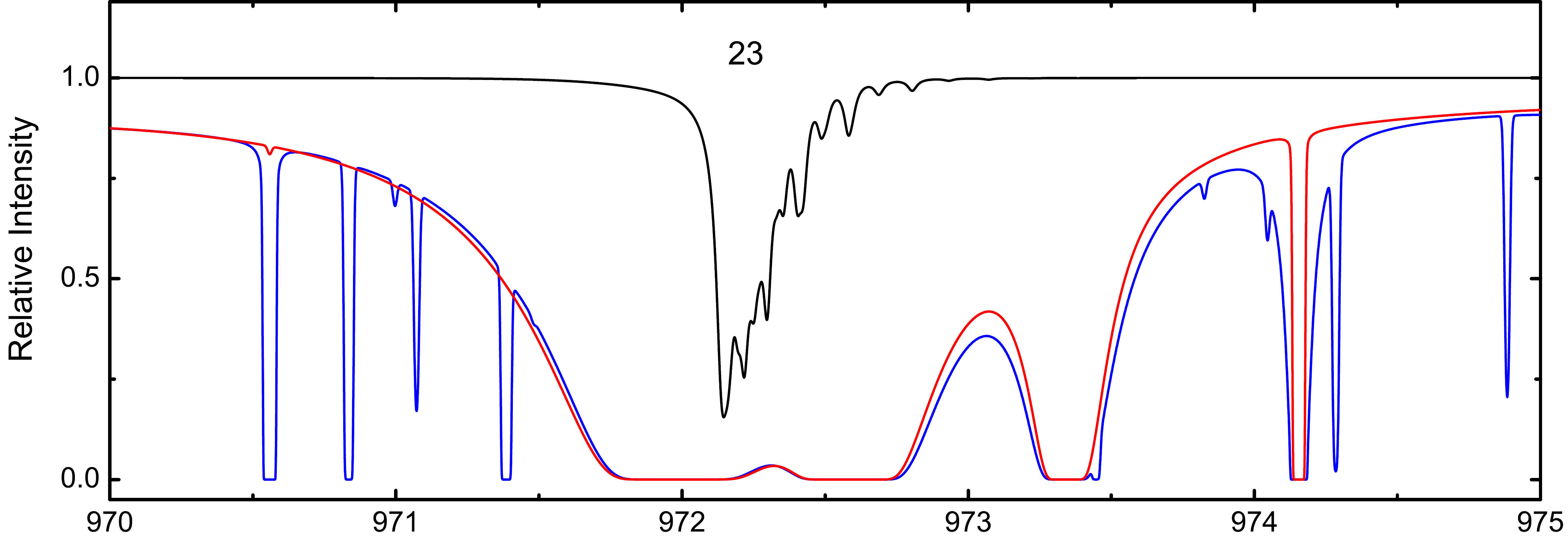}
\includegraphics[angle=0,width=0.9\hsize,height=5.4cm]{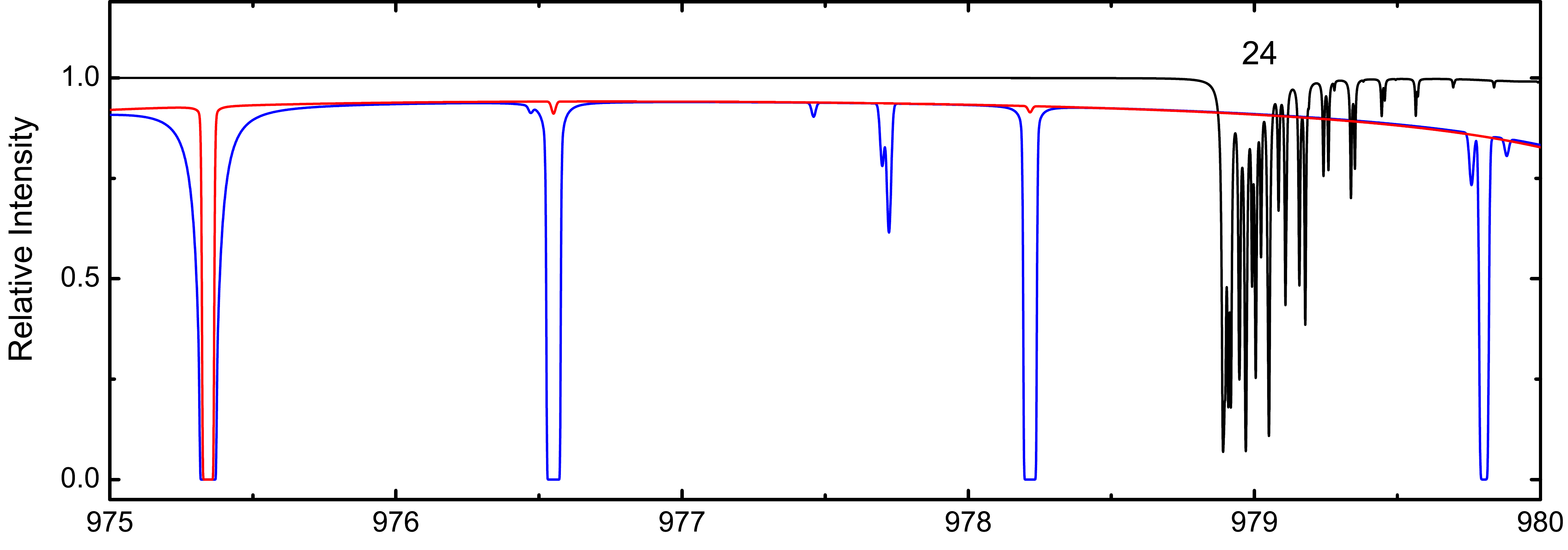}
\includegraphics[angle=0,width=0.9\hsize,height=5.4cm]{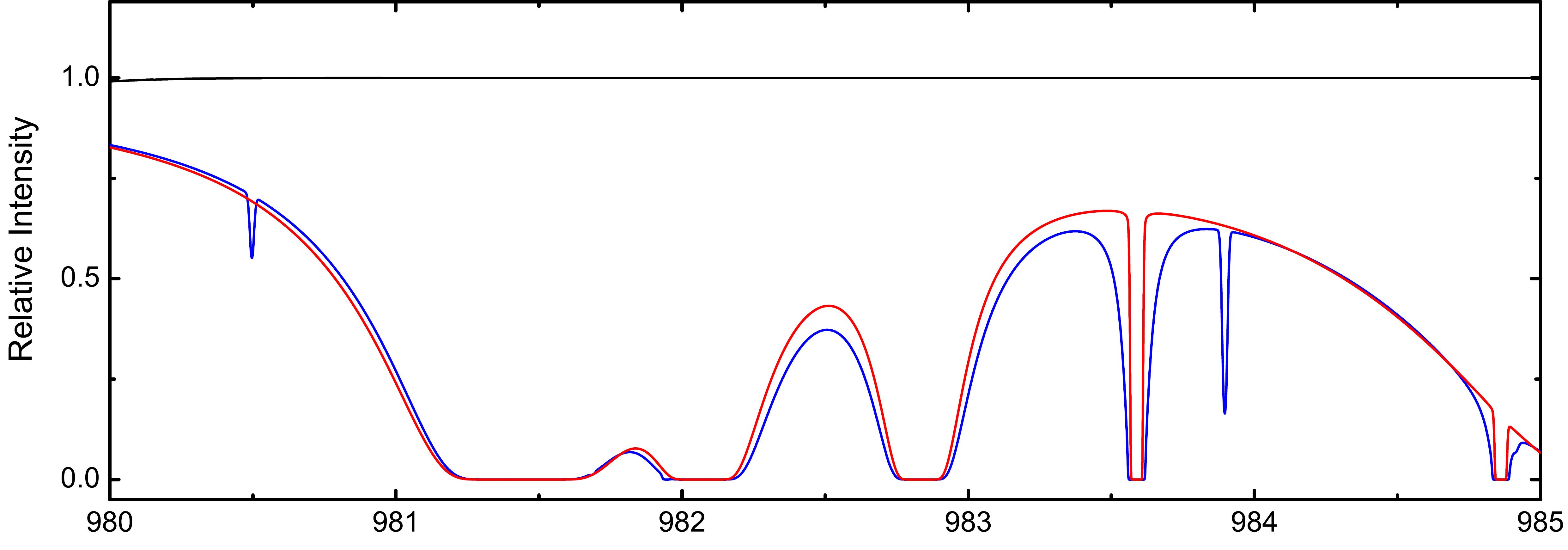}
\includegraphics[angle=0,width=0.9\hsize,height=5.4cm]{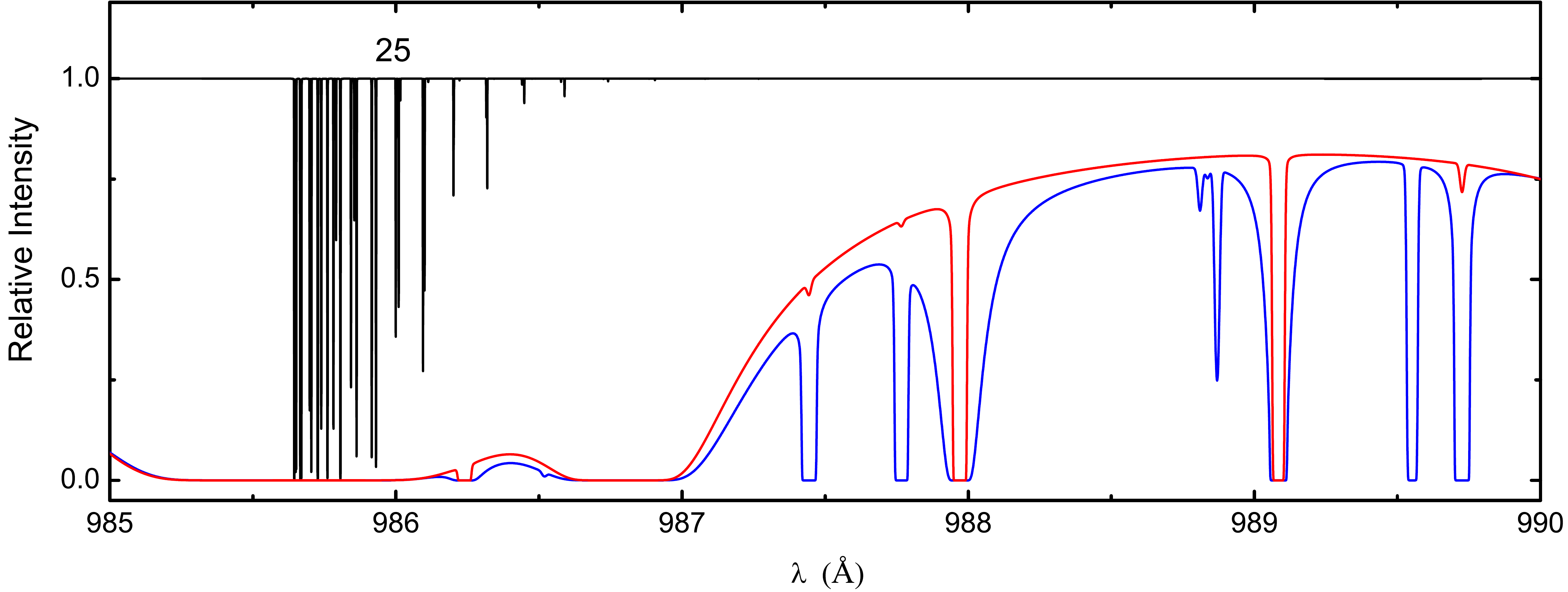}
\caption{As Fig.~\ref{fig:N2detail1}, but for 970--990 \AA  }
\label{fig:N2detail4}
\end{figure*}
\clearpage

\setcounter{figure}{4}
\begin{figure*}[htb]
\centering
\includegraphics[angle=0,width=0.9\hsize,height=5.4cm]{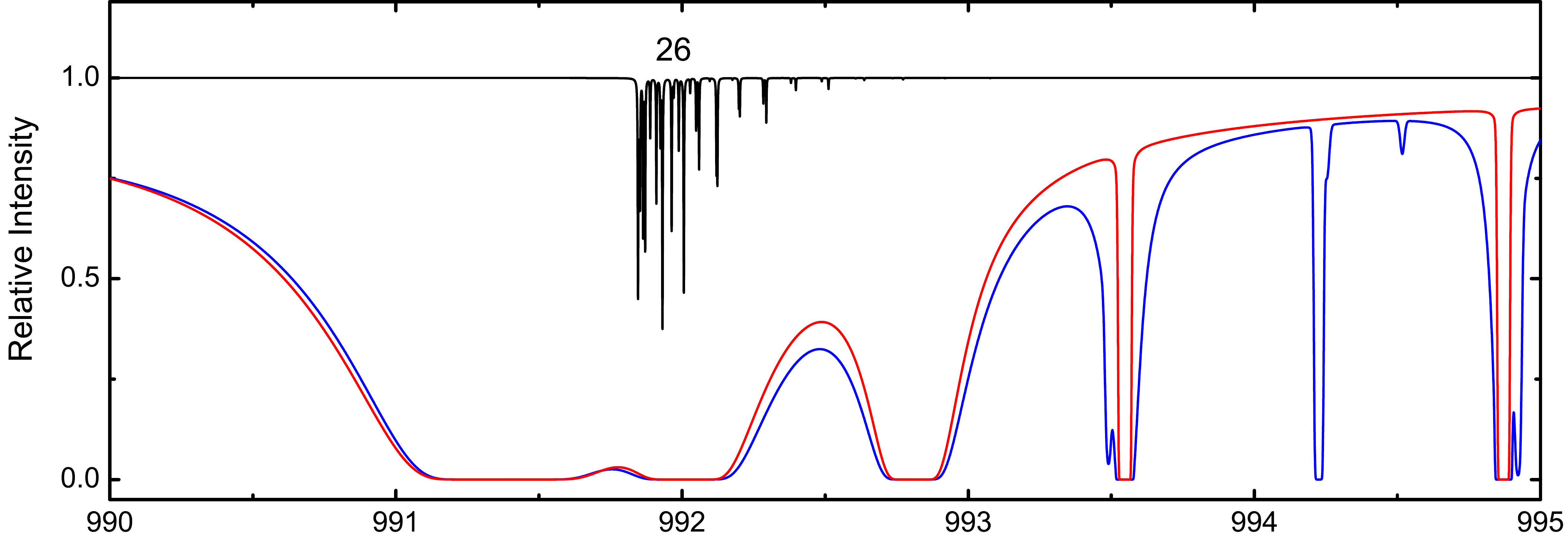}
\includegraphics[angle=0,width=0.9\hsize,height=5.4cm]{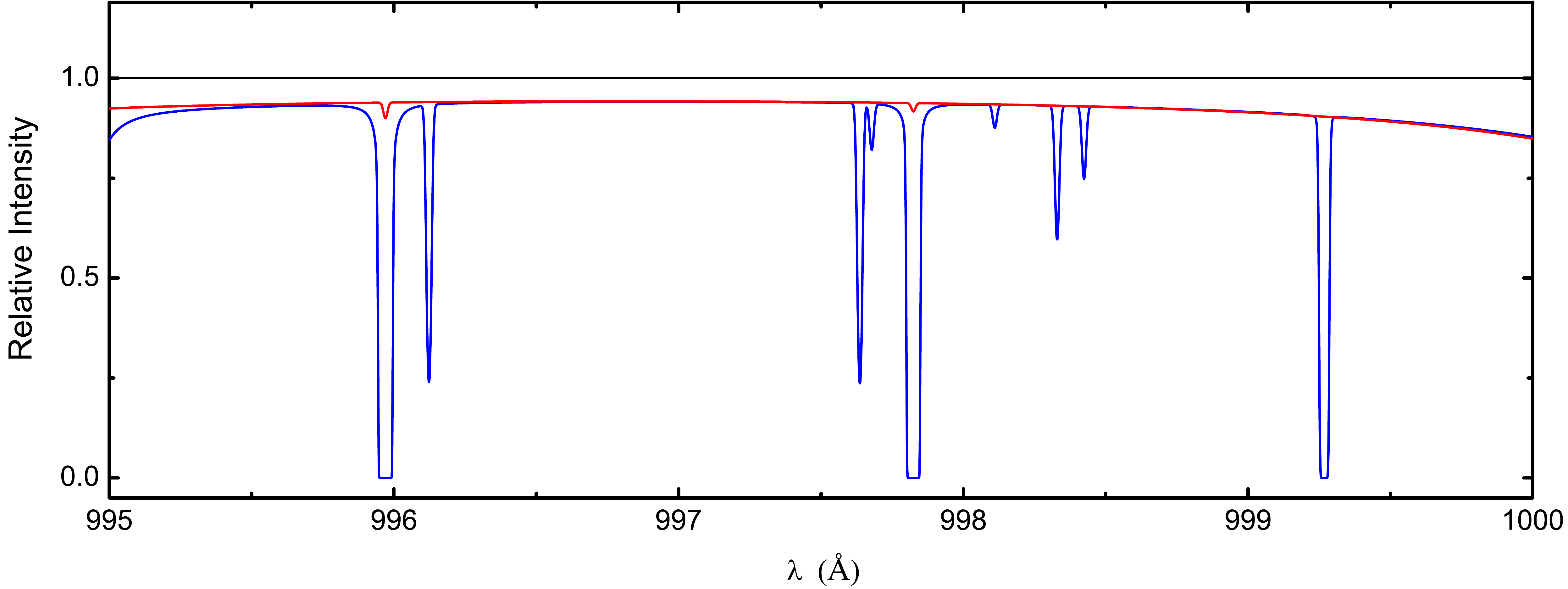}
\caption{As Fig.~\ref{fig:N2detail1}, but for 990--1000 \AA.}
\label{fig:N2detail5}
\end{figure*}
\clearpage
\end{appendix}
\end{document}